\documentclass[draftcls,12pt,onecolumn]{IEEEtran}
\usepackage{amsmath,amssymb,bm,cite,algorithm,algpseudocode,amsthm}
\usepackage{graphicx, xcolor}
\usepackage{bm}
\usepackage[printonlyused,withpage]{acronym}
\usepackage{booktabs}
\usepackage{tabularx}
\usepackage{subfigure}

\usepackage{color,graphicx,epsfig,psfrag}
\usepackage{pifont,enumerate,cases,algorithm,balance}
\usepackage{mathrsfs} 
\usepackage{multirow,multicol}
\usepackage{bm} 
\usepackage{verbatim}
\usepackage{algpseudocode}
\usepackage{caption}

\def\bb0{{\mathbb{0}}}

\def\ba{{\mathbf{a}}}
\def\bb{{\mathbf{b}}}

\def\bee{{\mathbf{e}}}

\def\bh{{\mathbf{h}}}

\def\bn{{\mathbf{n}}}

\def\bq{{\mathbf{q}}}
\def\br{{\mathbf{r}}}
\def\bs{{\mathbf{s}}}

\def\by{{\mathbf{y}}}
\def\bz{{\mathbf{z}}}
\def\b0{{\mathbf{0}}}

\def\bA{{\mathbf{A}}}

\def\bC{{\mathbf{C}}}
\def\bD{{\mathbf{D}}}

\def\bF{{\mathbf{F}}}
\def\bG{{\mathbf{G}}}
\def\bH{{\mathbf{H}}}
\def\bI{{\mathbf{I}}}

\def\bL{{\mathbf{L}}}

\def\bW{{\mathbf{W}}}
\def\bX{{\mathbf{X}}}
\def\bY{{\mathbf{Y}}}

\def\bbC{{\mathbb{C}}}

\def\bbE{{\mathbb{E}}}

\def\bbR{{\mathbb{R}}}

\def\cC{\mathcal{C}}

\def\cF{\mathcal{F}}

\def\cI{\mathcal{I}}

\def\cK{\mathcal{K}}

\def\cN{\mathcal{N}}
\def\cO{\mathcal{O}}

\def\cR{\mathcal{R}}

\def\sf0{{\mathsf{0}}}
 
\def\nn{\nonumber}

\DeclareMathOperator*{\argmax}{argmax}

\newtheorem{remark}{\underline{Remark}}

\def\teq{\triangleq} 
\def\j{\text{j}}

\def\T{\text{T}}
\def\R{\text{R}}

\def\olbA{{\overline{\bA}}}

\def\bGamma{\bm{\gamma}}

\def\bCR{\mathbf{C}_{\R}}

\def\Nt{{N_{\text{T}}}}
\def\Nr{{N_{\text{R}}}}

\def\Ns{{N_{\text{s}}}}  
\def\Lt{{L_{\text{T}}}}
\def\Lr{{L_{\text{R}}}}
\def\Kt{{K_{\text{T}}}}
\def\Kr{{K_{\text{R}}}}

\def\Nc{{N_{\text{c}}}}
\def\Nsa{{N_{\text{sa}}}}

\def\Ts{{T_{\text{s}}}} 
\def\lambdac{{\lambda_{\text{c}}}}
\def\fc{{f_{\text{c}}}}
\def\fs{{f_{\text{s}}}}

\def\dt{{d_{\text{T}}}}
\def\dr{{d_{\text{R}}}}

\def\vec{\mathrm{vec}}

\def\kron{\otimes} 

\def\diag{\text{diag}}

\def\cKcen{{\mathcal{K}}_{\text{cen}}}
\def\cKside{{\mathcal{K}}_{\text{side}}}

\def\alphal{{\alpha_{\ell}}}

\def\thetal{{\theta_{\ell}}}

\def\phil{{\phi_{\ell}}}

\def\taul{{\tau_{\ell}}}

\def\bPhi{{\mathbf{\Phi}}}

\usepackage{extarrows}
\usepackage{bm} 
\usepackage{geometry}
\usepackage{fancyhdr}
\begin{document}
\title{Hybrid mmWave MIMO Systems under Hardware Impairments and Beam Squint: Channel Model and Dictionary Learning-aided Configuration}
\author{Hongxiang Xie, Joan Palacios and Nuria Gonz\'{a}lez-Prelcic$^*$\thanks{$^*$Hongxiang Xie was with the Department of Electrical and Computer Engineering at the University of Texas at Austin when this work was performed. Nuria Gonz\'{a}lez-Prelcic and Joan Palacios are with the Electrical and Computer Engineering Department, North Carolina State University, USA.  
The corresponding author is  Nuria Gonz\'{a}lez-Prelcic: ngprelcic@ncsu.edu.}}
%\author{Hongxiang Xie, \emph{Graduate Student Member, IEEE}, Joan Palacios, \emph{Member, IEEE}, and Nuria Gonz\'{a}lez-Prelcic, \emph{Senior Member, IEEE}}  
\maketitle 
\vspace*{-1cm}
\begin{abstract}
Low overhead channel estimation based on compressive sensing (CS) has been widely investigated for hybrid wideband millimeter wave (mmWave) multiple-input multiple-output (MIMO) systems. The channel sparsifying dictionaries used in prior work are built from ideal array response vectors evaluated on discrete angles of arrival/departure. In addition, these dictionaries are assumed to be the same for all subcarriers, without considering the impacts of hardware impairments and beam squint. In this manuscript, we derive a general channel and signal model that explicitly incorporates the impacts of hardware impairments, practical pulse shaping functions,  and beam squint, overcoming the limitations of mmWave MIMO channel and signal models commonly used in previous work. Then, we propose a dictionary learning (DL) algorithm to obtain  the sparsifying dictionaries embedding hardware impairments, by considering the effect of beam squint without introducing it into the learning process.
We also design a novel CS channel estimation algorithm under beam squint and hardware impairments, where the channel structures at different subcarriers are exploited  to enable  channel parameter estimation with low complexity and high accuracy.   
Numerical results demonstrate the effectiveness of the proposed DL and channel estimation strategy when applied to realistic mmWave channels. 
\end{abstract}


%%%%%%%%%%%%%%%%%%%%%%%%%%%%%%%%%%%%%%%%%%%%%%%
\section{Introduction}\label{sec:intro}
%%%%%%%%%%%%%%%%%%%%%%%%%%%%%%%%%%%%%%%%%%%%%%%
The acquisition of channel state information (CSI) is crucial for mmWave link configuration, and challenging when operating with hybrid beamforming architectures.
To reduce the training overhead associated with CSI acquisition,
prior work has made full use of the sparse nature of mmWave channels in the angular or delay domains \cite{Vlachos2019,rodriguez2018frequency,HanLee:Two-stage-compressed-sensing:16, VenAlkPre:Channel-Estimation-for-Hybrid:17,LeeGilLee:Exploiting-spatial-sparsity:14}. 
Nevertheless, some relevant practical aspects have not been fully considered in previous compressive channel models and estimation algorithms: the beam squint effect, calibration errors,  and  hardware impairments.
Specifically, the channel sparsifying dictionaries used in prior work are typically assumed to be (overcomplete) discrete Fourier transform (DFT) matrices, or constructed from the ideal array response matrices (IARM) evaluated on discrete grids of quantized angles of arrivals and departures (AoAs/AoDs) \cite{rodriguez2018frequency,VenAlkPre:Channel-Estimation-for-Hybrid:17}. These assumptions are valid, however, only when the beam squint effect is negligible and no hardware impairments or calibration errors exist. In this paper, we show  that under hardware impairments such as mutual coupling or antenna separation disturbances,  the array response vectors will no longer be the Vandermonde vectors, and that different array response vectors should be considered at every frequency for channel modeling under beam squint.  
In other words, the assumptions and modeling of wideband mmWave MIMO channels in prior work are not valid, and therefore, the prior CSI acquisition strategies are not effective when beam squint and hardware impairments are considered.

The impact of beamsquint has been analyzed in prior work. As shown in   \cite{WanGaoJin:Spatial--and-Frequency-Wideband-Effects:18,BraSay:Wideband-communication-with:15,ChuHas:True-Time-Delay-Based-Multi-Beam-Arrays:13}, the time delay of the same data symbol across the antenna array aperture is non-negligible in the large-scale MIMO configurations and/or the wideband systems. Due to the spatial delay difference of each data symbol at different antennas, the array steering vectors will have different responses at each frequency, what leads to the beam squint effect. The work presented in \cite{WanGaoJin:Spatial--and-Frequency-Wideband-Effects:18} includes the derivation of a channel model for the large-scale MIMO system under beam squint, showing that the array response vectors for channel modeling have to be frequency-dependent. Aware of this beam squint impact,  \cite{rodriguez2018channel} and \cite{GonUtsCas:Hybrid-LISA-for-Wideband:18} also considered the frequency domain channel models by using explicit frequency-dependent array response vectors for different subcarriers.
Meanwhile, to avoid the impacts of beam squint on channel estimation at different subcarriers, \cite{wang2019block,jian2019angle,wang2019beam} proposed to estimate the channel parameters from the perspective of angle and delay domains, together with user scheduling to alleviate inter-user interference. The works in \cite{Tan2021a} and \cite{Tan2021b} consider 
a massive MIMO setting at mmWave or THz bands with single antenna users, and propose solutions for beam tracking and channel estimation, respectively, considering beam squint, but they do not introduce the filtering effect or hardware impairments in the channel model. Generally, these prior works on channel estimation/tracking under beam squint do not consider beam squint combined with filtering effects or other hardware imperfections.  In this manuscript, we will show  that beam squint not only leads to the frequency-dependence on array steering vectors, but also yields  additional distortions at different antennas across all subcarriers, especially on those at the band edge, when combined with the  filtering operations at the transceivers. The new channel model that results from these considerations has not been derived in  previous work.  The pioneering work in \cite{Schniter2014}  showed the need and impact of considering the pulse shaping function in the MIMO channel model, but beam squint was neglected.
In summary, prior work  did not consider the combined effect of beam squint and the filtering effects, and also neglected the impact of other hardware impairments such as calibration error, mutual coupling or antenna separation disturbances when developing the signal model and algorithms for a mmWave system operating with a  hybrid MIMO architecture.

The inclusion of hardware impairments on MIMO channel models  has been investigated in \cite{EbeEscBie:Investigations-on-antenna-array:16,ding2018dictionary,xie2020dictionary,palacios2019managing}. There are many different hardware impairments in the practical radio frequency (RF) chains, although 
three of them dominate the effects in the resulting channel. First, due to the manufacture and calibration errors, the antenna array will generate unexpected radiation patterns, including both gain and phase errors on each antenna element. Second, any perturbation on antenna locations or inter-element spacing between antenna elements will result in irregular linear arrays rather than perfect half-wavelength uniform linear arrays (ULAs). Finally, the antenna spacing disturbance also creates the mutual coupling effect between antenna elements. Taking into account all these hardware impairments, it is apparent that the aforementioned sparsifying dictionaries constructed from IARM evaluated at quantized angles are no longer the best choice for exploiting channel sparsity.  
Previous work has shown that
dictionary learning (DL) is an effective technique to capture the underlying structure of mmWave MIMO channels associated to specific types of sites \cite{WieWeiUts:Low-Rank-Approximations-for-Spatial:16} or with various hardware impairments \cite{ding2018dictionary, xie2020dictionary,palacios2019managing}. 
Specifically, \cite{WieWeiUts:Low-Rank-Approximations-for-Spatial:16} exploited the K-SVD algorithm to find a dictionary to represent a collection of observed channel realizations. Following this idea, \cite{ding2018dictionary} proposed  a joint uplink/downlink sparsifying DL algorithm for narrow band massive MIMO systems operating at lower frequencies. 
Our previous work \cite{xie2020dictionary,palacios2019managing} further investigated the DL and channel estimation strategy for hybrid wideband mmWave MIMO systems under low SNR conditions. Nevertheless, the beam squint effect was not incorporated in previous work, and it cannot be ignored under certain relationships between the carrier frequency and the bandwidth \cite{BraSay:Wideband-communication-with:15}. 

Motivated by these limitations, we propose a DL-based channel estimation strategy for hybrid mmWave MIMO systems under the impacts of both hardware impairments and beam squint.
The main contributions of this manuscript are summarized as follows: 
\begin{itemize} 
	\item We derive a general wideband mmWave MIMO channel model under both hardware impairments and beam squint. The new  model not only incorporates the hardware impairments of antenna spacing disturbances, gain/phase errors, and array mutual coupling, but also explicitly considers the impacts of combined pulse shaping, filtering and beam squint, showing the limitations of existing  MIMO channel models with beam squint and the associated channel estimation schemes. The derivation of the combined effect of beam squint and pulse shaping/filtering is of particular interest.  Previous literature does not provide an alternative channel model including all these effects. 
		\item We propose a DL algorithm for finding sparsifying dictionaries that embed hardware impairments, which fully exploits the channel properties at different subcarriers under beam squint. This algorithm does not learn the impact of beam squint but rather considers its model. Comparing to existing DL strategies which has to learn a general dictionary for both hardware impairments and beam squint, the newly proposed DL scheme enables better adaptation to the impacts of hardware impairments and facilitates the management of the beam squint impact at different subcarriers.  
	\item We design a novel orthogonal matching pursuit (OMP)-based algorithm for compressive channel estimation under beam squint, which exploits the simple structures of the new channel model at central subcarriers to obtain initial parameter estimates with low complexity. Then it compensates the additional distortions at side subcarriers induced by beam squint. In this way, the measurements at all subcarriers can be used to achieve higher parameter estimation accuracy with lower complexity.       
	\item We evaluate the proposed DL and channel estimation algorithms via numerical simulations. Results show that the training overhead of channel estimation with learned dictionaries can be significantly reduced compared to traditional dictionaries without considering hardware impairment or beam squint. This validates the developed channel models and corroborates the effectiveness of the proposed DL algorithms for hybrid wideband mmWave MIMO systems under both hardware impairments and beam squint effects.      
\end{itemize}  

\textbf{Notations:} Vectors and matrices are denoted by boldface small and
capital letters;  the transpose, conjugate, Hermitian (conjugate transpose),
inverse, and pseudo-inverse of the matrix $\bA$ are denoted by 
$\bA^T$, $\olbA$, $\bA^*$, $\bA^{-1}$ and $\bA^\dag$;  
${\bI_M}$ is an $M\times M$ identity matrix;  
$[\ba]_{n}$ denotes the $n$-th element of $\ba$ and  
$[\bA]_{:,j}$ denotes the $j$-th column vector of $\bA$; 
$[\bA]_{m,n}$ denotes the $(m,n)$-th element of $\bA$;
$\teq$ represents new definitions; 
$\cI(N)\teq \{0,1,\ldots,N-1\}$ denotes the index set of cardinality $N$;
$\bbC$ and $\bbR$ denote the sets of complex and real numbers; 
$\text{tr}\{\bA\}$ is the trace of $\bA$;   
$\diag\{\ba\}$ denotes a diagonal matrix with its diagonal elements given in $\ba$ and
$\diag\{\bA\}$ formulates a vector by extracting the diagonal elements of $\bA$; 
$\kron$, $\odot$ and $\star$ denote the Kronecker, Hadamard and Khatri-Rao product between vectors/matrices;
$\j =\sqrt{-1}$ denotes the imaginary unit;
$|\cK|$ denotes the cardinality of a set;
$\lfloor x\rfloor$ denotes the largest integer less than or equal to $x$ and 
$\lceil x\rceil$ denotes the smallest integer great than or equal to $x$;
$\cF(\cdot)[f]$ denotes the Fourier transform evaluated at frequency $f$;
and $\|\ba\|_0$ denotes the $\ell_0$ norm of vector, i.e., the number of nonzero elements of $\ba$. 
 
\section{New Signal and Channel Models Incorporating Hardware Impairments and Beam Squint}

In this section, we derive the new general time domain signal model and frequency domain channel model for the hybrid mmWave MIMO systems under the impact of both hardware impairments and beam squint. We will prove that the channel models with beam squint  assumed in previous works \cite{WanGaoJin:Spatial--and-Frequency-Wideband-Effects:18,GonUtsCas:Hybrid-LISA-for-Wideband:18,rodriguez2018channel,wang2019block,jian2019angle,wang2019beam} are not complete or valid in some cases.  

We consider a fully connected hybrid mmWave MIMO system.%as illustrated in \figref{fig:systemmodel}. 
The transmitter (TX) is equipped with $\Nt$ antennas and $\Lt$ RF chains, and the receiver (RX) has $\Nr$ antennas and $\Lr $ RF chains. 
The channel between the TX and the RX is frequency-selective. Pulse shaped orthogonal frequency division multiplexing (OFDM) with $K$ subcarriers is considered to simultaneously transmit $\Ns~(\leq \min(\Lt,\Lr))$ data streams. The system sampling period is denoted by $\Ts$.   
We also consider a pulse shaping function with a roll-off factor $\beta$ and an overall system bandwidth of $(1+\beta)/\Ts$. The center carrier frequency and wavelength are denoted by $\fc$ and $\lambdac=c/\fc$ (with $c$ the speed of light). 
We use the index $k$ to denote the frequency domain subcarriers, with 
\begin{align}\label{equ:fk} 
f_k &= f_\text{c} + \Delta f_k = f_\text{c} - \frac{1}{2\Ts} + \frac{k}{K\Ts}, \quad \text{for}\ k=0,\ldots,K-1.
\end{align} 
the corresponding subcarrier frequency value \cite{tan2004reduced}.

Let us start by defining a model for practical antenna arrays. %It is common in the literature to assume that all the antennas are arranged in the form of ULAs at both TX and RX, whose expected ideal antenna spacings are denoted by $\dt$ and $\dr$, respectively. In this work, we also assume the linear architecture for TX and RX arrays, but take into consideration some practical imperfections.
As shown in \cite{friedlander1991direction, EbeEscBie:Investigations-on-antenna-array:16,ding2018dictionary,xie2020dictionary,stavropoulos2000array}, various impairments exist in practical implementations of antenna arrays, such as the gain/phase errors on each antenna element, the errors at each antenna element location, and the coupling effects between antenna elements.  Specifically, we denote $\bCR\in\bbC^{\Nr\times\Nr}$ as the symmetric mutual coupling matrix for the RX antenna array, representing the unwanted interchange of energy between elements in the arrays, and denote $\bm\gamma_{\R}\in\mathbb{C}^{N_{\R}\times 1}$ with $[\bm\gamma_{\R}]_{n_{\R}}=g_{\R,n_{\R}}e^{\j\nu_{\R,n_{\R}}}$ as the antenna gain and phase errors, in which $g_{\R,n_{\R}}$ and $\nu_{\R,n_{\R}}$ are the receive gain error normalized to a reference amplitude, and the additional receive phase error for the $n_{\R}$-th antenna element. Moreover, let $\bm\epsilon_{\R}\in\bbR^{\Nr \times 1}$ with $[\bm\epsilon_{\R}]_{n_\R} = \epsilon_{\R,n_\R}$ be the vector of antenna location errors at all RX antenna elements, where the location error of the first antenna element is normalized to $\epsilon_{\R,0} =0$.
The variables defining the hardware impairments for the TX antenna array are defined in an analogous way. 
%\begin{figure}[t!]
%	\centering 
%	\includegraphics[width=160mm]{./figures-old/systemmodel.pdf} 
%	\caption{System model and architecture of a hybrid mmWave MIMO system with hardware impairments on antennas.}\label{fig:systemmodel}
%\end{figure}  

We define now the signal model for the transmitted signal during training. To sound the channel, the TX sends $Q$ symbols over a $\Lt\times 1$ signal vector ${\bf s}(t)$ at time instant $t$, where the $l_{\T}$-th ($l_{\T}=1,\ldots,\Lt$) element of the complex exponential representation of the transmitted signal ${\bf s}(t)$ is defined as
\begin{align}
	s_{l_{\T}}(t)\teq [{\bf s}(t)]_{l_{\T}}  = \sum_{q=1}^{Q} [{\bf F}_{\rm BB}{\bf S}]_{l_{\T},q} p_{\rm T}(t-(q-1)\Ts)e^{j2\pi f_{\rm c}t},
	\label{eq:TXsignal}
\end{align}
where $p_{\rm T}(.)$ is the transmit pulse, $e^{j2\pi f_{\rm c}t}$ models the RF upconversion stage, ${\bf S}\in\mathbb{C}^{\Lt\times Q}$ collects all the time domain signal symbols to be transmitted and ${\bf F}_{\rm BB}\in\mathbb{C}^{\Lt\times\Lt}$ is the digital precoder.
Moreover, ${\bf s}(t)$ satisfies $\bbE\{{\bf s}(t){\bf s}(t)^*\}=\frac{P_{\T}}{\Lt}\bI_{\Lt}$, with $P_{\T}$ the transmit power constraint. Using an analog precoder $\bF_{\rm RF} \in\mathbb{C}^{\Nt \times \Lt}$, the ideal transmitted signal at the $n_{\rm T}$-th antenna element,  $n_{\rm T}=1,\ldots,\Nt$, can be computed as
\begin{align}
i_{n_{\rm T}}(t) = \sum_{l_{\T}=1}^{\Lt} [\bF_{\rm RF}]_{n_{\rm T}, l_{\T} } [{\bf s}(t)]_{l_{\T}}.
\end{align}
Due to hardware imperfections, the signal at the  ${n_\T}$-th transmit antenna  is multiplied by a phase and amplitude perturbation $[\bGamma_{\T}]_{n_\T}^*$, and then leaks into the signal at the ${n'_\T}$-th transmit antenna due to mutual coupling by a factor $[{\bf C}_\T]_{n_\T, n'_\T}^*$, leading to the expression of the actual transmitted signal
\begin{align}\label{equ:r_nT}
x_{n_{\rm T}}(t) = \sum_{n'_{\rm T}=1}^{\Nt} [{\bf C}_\T]_{n'_\T, n_\T}^*[\bGamma_{\T}]_{n'_\T}^*i_{n'_{\rm T}}(t) = \sum_{n'_{\rm T}=1}^{\Nt} [\left(\diag(\bGamma_{\T}){\bf C}_\T\right)^*]_{n_{\rm T}, n'_{\rm T}}i_{n'_{\rm T}}(t).
\end{align}
Assuming a passband geometric channel model with $L$ paths, the ideal received signal at the $n_{\rm R}$-th antenna element, $n_{\rm R}=1,\ldots,\Nr$, can  be written as
\begin{align} 
i_{n_{\rm R}}(t) = \sum_{\ell=1}^L \sum_{n_{\T}=1}^{\Nt} \alpha_l^{\rm P} \cdot x_{n_{\rm T}}(t-\tau_{\ell,n_{\R},n_{\T}})  + z_{n_{\rm R}}(t),
\end{align}
where $z_{n_{\rm R}}(t)$ is the noise term, $\alpha_l^{\rm P}\in \bbC$ is the path gain of the passband channel including phase changes, and $\tau_{\ell,n_{\R},n_{\T}}$ is the delay of $\ell$-th path between the $n_{\R}$-th receive antenna and the $n_{\T}$-th transmit antenna. Similarly to what happens at the transmit array, owing to hardware imperfections, the signal at each receive antenna ${n_\R}$ leaks into the signal at each antenna ${n'_\R}$ through mutual coupling by a factor $[{\bf C}_\R]_{n'_\R, n_\R}$, and then receives a gain/phase weight $[\bGamma_{\R}]_{n'_\R}$ to generate the actual receive signal
\begin{align}
y_{n_{\rm R}}(t) = \sum_{n'_{\rm R}=1}^{\Nr} [\bGamma_{\R}]_{n_\T}[{\bf C}_\R]_{n_\R, n'_\R} i_{n'_{\rm R}}(t) = \sum_{n'_{\rm R}=1}^{\Nr} [\diag(\bGamma_{\R}){\bf C}_\R]_{n_\R, n'_\R} i_{n'_{\rm R}}(t).
\end{align} 
After an analog combining stage modeled by the matrix ${\bW}_{\rm RF}\in\mathbb{C}^{N_{\rm R}\times L_{\rm R}}$, we obtain the received signal vector $\bm y'(t)\in \bbC^{\Lr\times 1}$, with the $l_{\R}$-th element given as  
\begin{align}
[\bm y'(t)]_{l_{\rm R}} = \sum_{n_{\rm R}=1}^{\Nr} [{\bf W}_{\rm RF}]_{n_{\rm R}, l_{\rm R}}^*y_{n_{\rm R}}(t).
\end{align}

To obtain the downconverted signal we get rid of the complex representation of the signal and multiply by $2e^{-j2\pi f_{\rm c}t}$ to get $[\bm y'']_{l'_{\rm R}}(t) = ([\bm y']_{l'_{\rm R}}(t) + [\bm y']_{l'_{\rm R}}^*(t))e^{-j2\pi f_{\rm c}t}$.
After matched filtering and sampling with sampling period $T_{\rm s}$, we have
\begin{align}
\big[{\bm  y}_{\rm BB}[d]\big]_{{l_{\rm R}}} = (p_{\rm R} * [\bm y'']_{l'_{\rm R}})((d-1)\Ts), \ \text{for}\ d=0,1,\ldots
\end{align}
where $p_{\rm R}$ is the matched filter.
Since the bandwidths of $p_{\rm T}$ and $p_{\rm R}$, namely $B_{\rm T}$ and $B_{\rm R}$, satisfy $B_{\rm T}+B_{\rm R} \leq 2f_{\rm c}$ and  $[\bm y']_{l'_{\rm R}}^*(t)e^{-j2\pi f_{\rm c}t}$ is centered at $-2f_{\rm c}$, the convolution between the matched filter and  the conjugate signal is zero, i.e.  $p_{\rm R}(t) * ([\bm y']_{l'_{\rm R}}^*(t)e^{-j2\pi f_{\rm c}t}) = 0$.
After the digital combining stage represented by ${\bf W}_{\rm BB}$, the discrete time received signal can be written as  
\begin{align}
\big[{\bm  y}[d]\big]_{{l_{\rm R}}} = \sum_{l'_{\rm R}}[{\bf W}_{\rm BB}]_{l'_{\rm R}, l_{\rm R}}^*\big[{\bm  y}_{\rm BB}[d]\big]_{{l_{\rm R}}}, \ \text{for}\ d=0,1,\ldots
\end{align}
By developing the previous steps, we reach an equivalent expression for the received signal
\begin{align} \label{equ:yd}
\big[{\bm  y}[d]\big]_{{l_{\rm R}}} \hspace*{-2mm}= \hspace{-5mm}\sum_{\ell, n_{\rm R}, n_{\rm T}, l_{\rm T}} \hspace*{-4mm}\alpha_l^{\rm P} [\breve{\bf W}]_{n_{\rm R}, l_{\rm R}}^*[\breve{\bf F}]_{n_{\rm T}, l_{\rm T}} \sum_{q=1}^Qp((d-q)T_{\rm s}- {\tau}_{\ell, n_{\rm R}, n_{\rm T}})e^{-j2\pi f_{\rm c}{\tau}_{\ell, n_{\rm R}, n_{\rm T}}} [\bs[q]]_{{l_{\rm T}}} \hspace*{-1mm}+ [\breve{\bf W}]_{n_{\rm R}, l_{\rm R}}^*z_{n_{\rm R}}(dT_{\rm s}), 
\end{align} 
where $\breve{\bf W} \teq \left(\diag(\bGamma_{\R}){\bf C}_\R\right)^*{\bf W}$ and $\breve{\bf F} \teq \left(\diag(\bGamma_{\T}){\bf C}_\T\right)^*{\bf F}$, being ${\bf W} = {\bf W}_{\rm RF}{\bf W}_{\rm BB}$ and ${\bf F} = {\bf F}_{\rm RF}{\bf F}_{\rm BB}$  the hybrid combiner and precoder, respectively, while $p = p_{\rm R}*p_{\rm T}$ combines the effects of transmit and receive filters.
\begin{figure}[t]
	\centering
	\subfigure[Propagation of the $\ell$-th path between RX and TX with AoA $\phil$ and AoD $\thetal$.]{
		\includegraphics[width=0.48\textwidth]{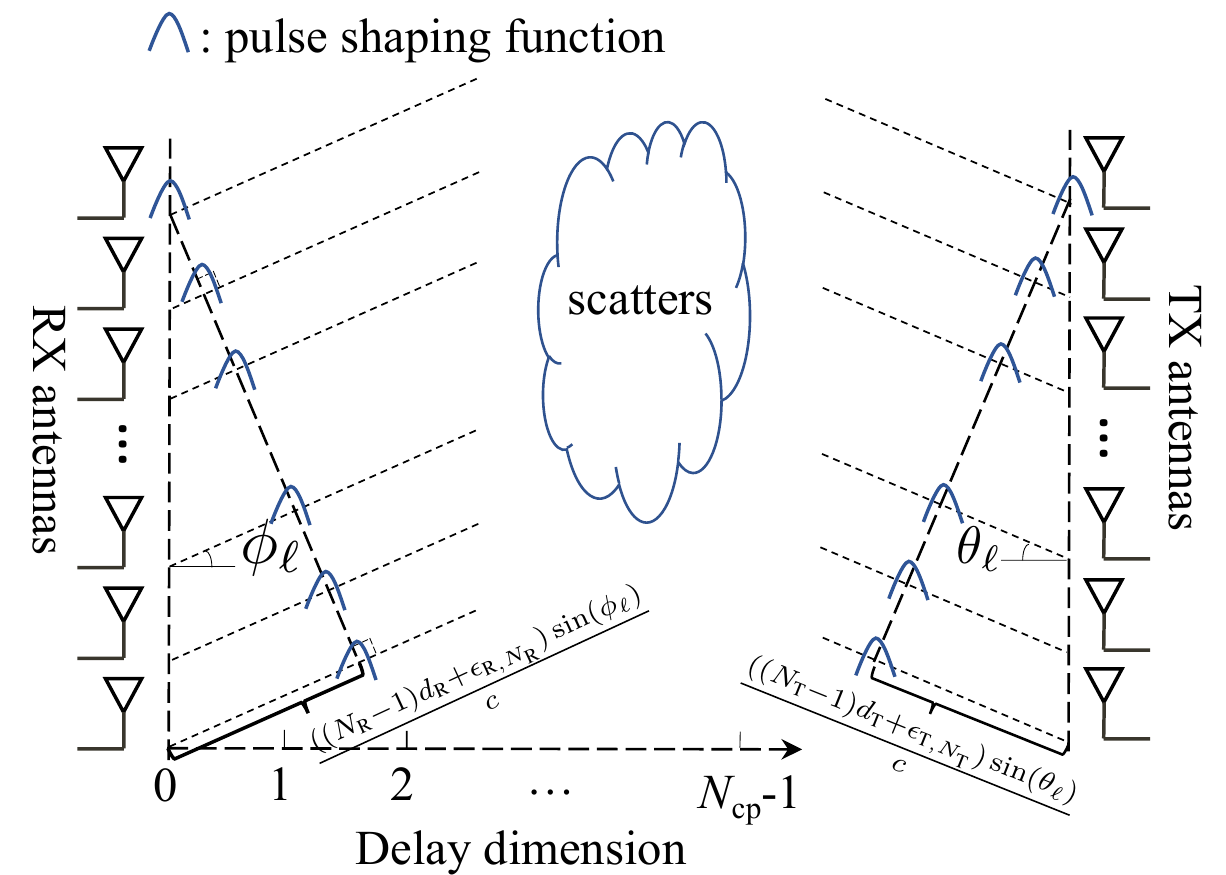}\label{fig:spatial-temporal3}}
	\quad 
	\subfigure[Received signals at RX antennas, where the delay of each path varies across array geometry.  ]{
		\includegraphics[width=0.45\textwidth]{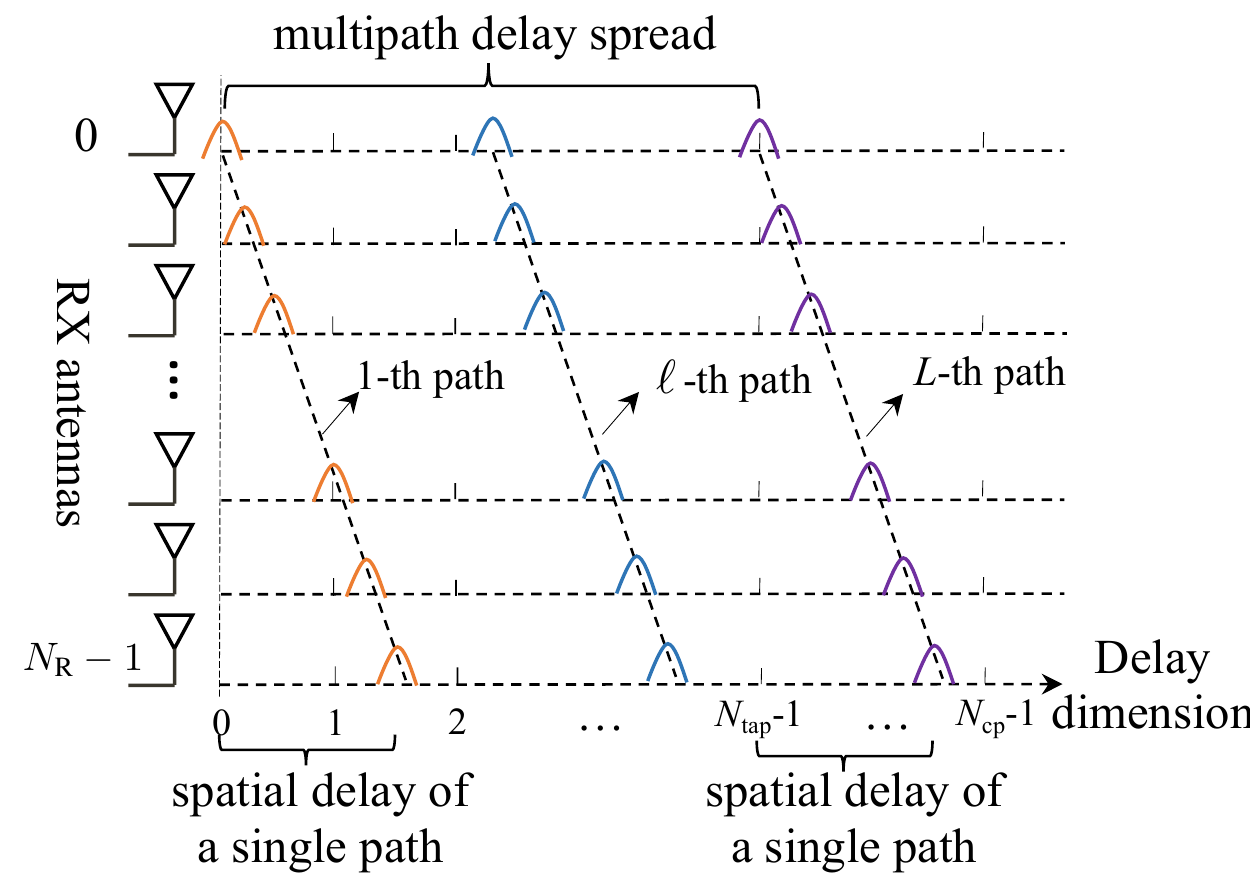}\label{fig:spatial-temporal4}} 
	\caption{Illustration of the spatial wideband effect (from the perspective of time domain) or the beam squint effect (from the perspective of frequency domain) under large-scale antenna array regimes.} \label{fig:spatial-temporal0} 
\end{figure} 
Next, following the derivation in the Appendix I, the DFT of \eqref{equ:yd} is computed as
\begingroup 
\allowdisplaybreaks 
\begin{align} \label{equ:yk_predef}
[{\bf y}[k]]_{{l_{\rm R}}} %&= \sum_{d=0}^{K-1} \big[{\bm  y}[d]\big]_{{l_{\rm R}}} e^{-\j\frac{2\pi dk}{K}} \notag \\
= \sum_{\ell, n_{\rm R}, n_{\rm T}, l_{\rm T}} \hspace*{-2mm}\alpha_l^{\rm P} [\breve{\bf W}]_{n_{\rm R}, l_{\rm R}}^*[\breve{\bf F}]_{n_{\rm T}, l_{\rm T}} g(k,  {\tau}_{\ell, n_{\rm R}, n_{\rm T}})e^{-\j 2\pi(f_{\rm c} + \Delta f_k) {\tau}_{\ell, n_{\rm R}, n_{\rm T}}}[ \bs[k]]_{{l_{\rm T}}} + [\breve{\bf W}]_{n_{\rm R}, l_{\rm R}}^*[ \bz[k]]_{n_{{\rm R}} },
\end{align}
\endgroup
where $g(k,  {\tau}_{\ell, n_{\rm R}, n_{\rm T}})$ is defined in \eqref{equ:g_k_tau} in the Appendix I, representing the frequency response at the $k$-th subcarrier of  the combined filter $p$ for the $\ell$-th path between the $n_{\R}$-th RX antenna and the $n_{\T}$-th TX antenna. Moreover, $[ \bs[k]]_{{l_{\rm T}}}$ is the frequency domain version of the transmitted signal defined in \eqref{eq:TXsignal}, while  $[\bz[k]]_{n_{\rm R}} \teq \sum_{d=0}^{K-1}z_{n_{\rm R}}(dT_{\rm s})   e^{-\j\frac{2\pi dk}{K}} $ is the noise term in the  frequency domain. 
Regarding the delay $ {\tau}_{\ell, n_{\rm R}, n_{\rm T}}$, 
as shown in Fig. \ref{fig:spatial-temporal0}, it can be expressed as\begin{align}
 {\tau}_{\ell, n_{\rm R}, n_{\rm T}} = \taul+(n_{\rm R}d_{\R}+\epsilon_{\R,n_{\rm R}})\sin(\phil)/c-(n_{\rm T}d_{\T}+\epsilon_{\T,n_{\rm T}})\sin(\thetal)/c.
\end{align}
Now, by defining the frequency-dependent array steering vectors 
\begin{align} \label{equ:aR_all}
\breve{\ba}_{\R,k}(\phi)=\diag(\bGamma_{\R}){\bf C}_\R\check{\ba}_{\R,k}(\phi),\\
\breve{\ba}_{\T,k}(\theta)=\diag(\bGamma_{\T}){\bf C}_\T\check{\ba}_{\T,k}(\theta),
\end{align} 
with 
\vspace*{-8mm}
\begin{align} \label{equ:baRT}
[\check{\ba}_{\R,k}(\phi)]_n=\frac{1}{\sqrt{N_{\rm R}}}e^{-\j2\pi f_k(n\dr+\epsilon_{\R,n})\sin(\phi)/c}\\
[\check{\ba}_{\T,k}(\theta)]_n=\frac{1}{\sqrt{N_{\rm T}}}e^{-\j2\pi f_k(n\dt+\epsilon_{\T,n})\sin(\theta)/c},
\end{align}
denoting the distortion matrix by
\begin{align}\label{equ:dk_tau}
[{\bf G}_k(\taul, \phil, \thetal)]_{n_{\rm R}, n_{\rm T}} = g(k,  {\tau}_{\ell, n_{\rm R}, n_{\rm T}}),
\end{align}
and defining the quivalent complex gain $\alpha_l = \alpha_l^{\rm P}e^{-\j2\pi f_{\rm c}\tau_l}$, we can rewrite \eqref{equ:yk_predef} as\begin{align} \label{equ:yk_def}
{\bf y}[k] = {\bf W}^{*}\sum_{\ell=1}^L\left( \alphal e^{-\j 2\pi\Delta f_k\taul} {\bf G}_k(\taul, \phil, \thetal)\odot(\breve{\ba}_{\R,k}(\phil) \breve{\ba}_{\T,k}^{*}(\thetal))\right) {\bf F}\hat{\bf s}[k] +  \breve{\bf W}^{*} {\bf z}[k].
\end{align}
Based on this expression, we can define the general frequency domain channel matrix as
\begin{align} \label{equ:H_def}
\bH[k] = \sum_{\ell=1}^L \alphal e^{-\j 2\pi\Delta f_k\taul} {\bf G}_k(\taul, \phil, \thetal)\odot(\breve{\ba}_{\R,k}(\phil) \breve{\ba}_{\T,k}^{*}(\thetal)),
\end{align}
and then, the final expression of the received signal under these definitions becomes
\begin{align} \label{equ:yk_H_def}
{\bf y}[k] = {\bf W}^{*}\bH[k] {\bf F}{\bf s}[k] +  \breve{\bf W}^{*}\bz[k].
\end{align}  
Note that the equivalent combiner that impacts the noise term  includes the hardware impairments as well, and thereby the noise covariance matrix is $\bC_{\bz[k]}= \sigma^2 \breve{\bf W}^{*}\breve{\bf W}$.

As per the definition of $g(k, \tau)$ in the Appendix I, i.e.,
	\begin{align}
	g(k, \tau) = \left\lbrace
	\begin{array}{cl}
	\frac{\mathcal{F}(p)[\Delta f_k]}{T_{\rm s}}+\frac{\mathcal{F}(p)[\Delta f_k-1/T_{\rm s}]}{T_{\rm s}}e^{\j2\pi\tau/T_{\rm s}} & \text{if}\ \ 2\Delta f_k > (1-\beta)/T_{\rm s}, \\
	\frac{\mathcal{F}(p)[\Delta f_k]}{T_{\rm s}} & \text{if} \ \  2|\Delta f_k| \leq (1-\beta)/T_{\rm s}, \\
	\frac{\mathcal{F}(p)[\Delta f_k]}{T_{\rm s}}+\frac{\mathcal{F}(p)[\Delta f_k+1/T_{\rm s}]}{T_{\rm s}}e^{-\j2\pi\tau/T_{\rm s}} & \text{if}\ \ 2\Delta f_k < -(1-\beta)/T_{\rm s},
	\end{array}
	\right.
	\end{align}
	where $\cF(p)[\Delta f_k]$ is the Fourier transform of the combined filter $p(t)$ evaluated for the frequency difference $\Delta f_k$, the general channel matrix in \eqref{equ:H_def} can be rewritten as 
	\begin{align}\label{equ:Hk_two_cases}
	\hspace*{-3mm}\bH[k] =  
	\begin{cases}
	  \sum_{\ell=1}^L \alphal e^{-\j 2\pi\Delta f_k\taul} {\bf G}_k(\taul, \phil, \thetal)\odot(\breve{\ba}_{\R,k}(\phil) \breve{\ba}_{\T,k}^{*}(\thetal)), \quad \forall  |\Delta f_k|> (1-\beta)/2\Ts \\
	  \sum_{\ell=1}^L \alphal \frac{\mathcal{F}(p)[\Delta f_k]}{\Ts}e^{-\j 2\pi\Delta f_k\taul}  \breve{\ba}_{\R,k}(\phil) \breve{\ba}_{\T,k}^{*}(\thetal), \quad \forall  |\Delta f_k|\leq (1-\beta)/2\Ts 
	\end{cases} 
	\end{align}
	where the second case is due to the fact that when $|\Delta f_k|\leq (1-\beta)/2\Ts $, the factor  $ g(k,  {\tau}_{\ell, n_{\rm R}, n_{\rm T}}) = \frac{\mathcal{F}(p)[\Delta f_k]}{\Ts}$ is independent of the delay $ {\tau}_{\ell, n_{\rm R}, n_{\rm T}}$ at different antenna indices, and thus enables the simplification. 
	Note that \eqref{equ:Hk_two_cases} is valid for any  pulse shaping function (as shown in the derivation in Appendix I), and therefore, this does not introduce any new assumption on the channel model. 
	
\begin{figure}[t]
	\centering 
	\includegraphics[width=120mm]{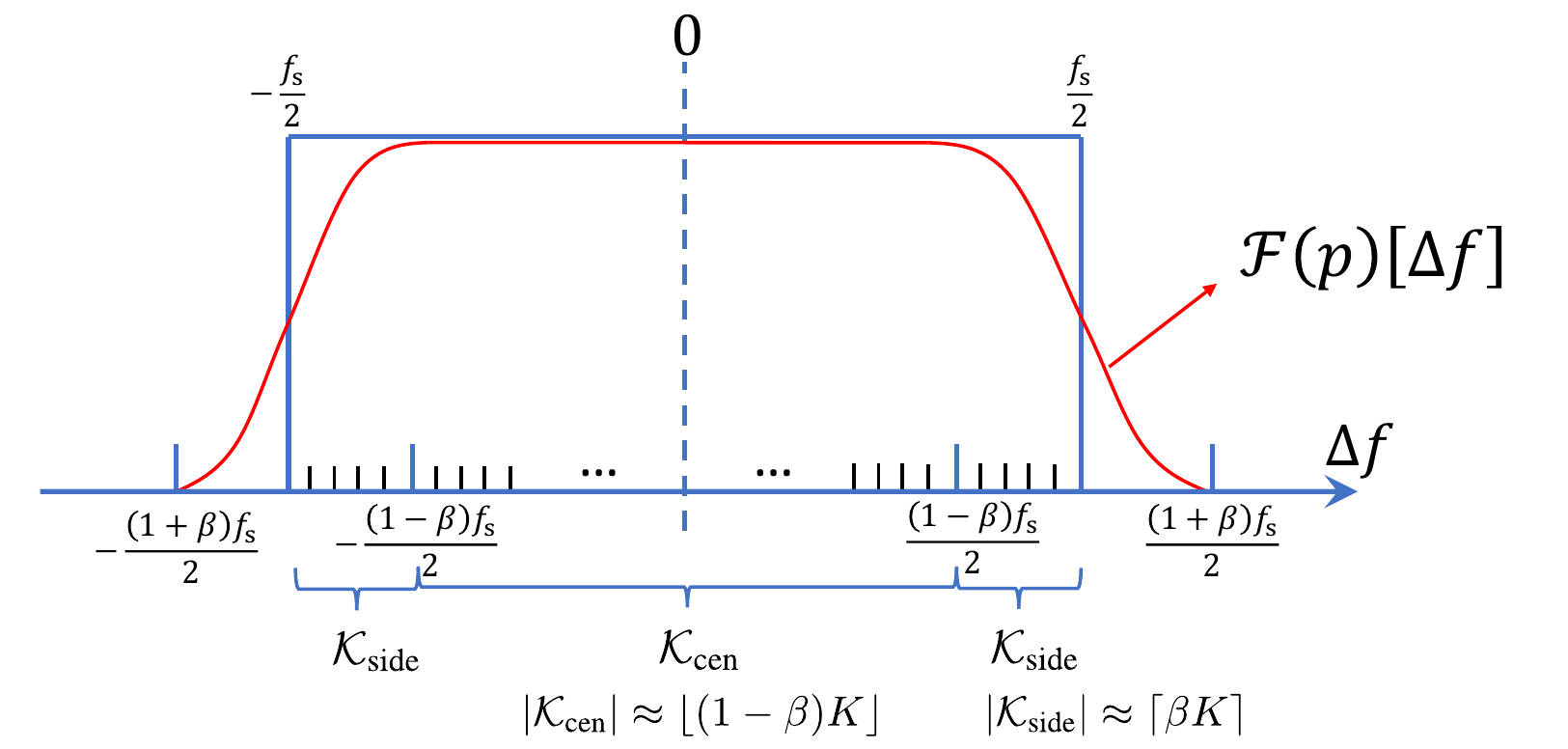} 
	\caption{Frequency response of the pulse shaping function with roll-off factor $\beta$, center frequency $\fc$ and $\fs=1/\Ts$. For OFDM systems with $K$ subcarriers, there are approximately $\lfloor(1-\beta)K\rfloor$ central subcarriers inside $\cKcen$, and $\lceil\beta K\rceil$ side subcarriers inside $\cKside$. }\label{fig:subcarrier_grouping}
\end{figure}

\begin{remark}
	We have derived the new general model for MIMO channels under both hardware impairments and beam squint. This new model shows that the beam squint not only induces the frequency-dependence on array steering vectors, but also yields   additional distortions at different antennas across all subcarriers, especially on side subcarriers. Comparing \eqref{equ:Hk_two_cases} to the existing MIMO channel models with beam squint in \cite{WanGaoJin:Spatial--and-Frequency-Wideband-Effects:18,GonUtsCas:Hybrid-LISA-for-Wideband:18,rodriguez2018channel,wang2019block,jian2019angle,wang2019beam}, it is obvious that those models only considered the second case in \eqref{equ:Hk_two_cases}, and assumed the same expression for all subcarriers.  In other words, the prior work only considered the frequency-dependence impact of beam squint, but ignored the additional distortions.
	This is due to the fact that the model was derived in the continuous time domain, and did not take into account the impact of the extra bandwidth of the combined filter (as shown in Appendix I).  
	For ease of subsequent exposition, we will denote the set of central subcarriers by $\cKcen =\{k \big| |\Delta f_k|\leq (1-\beta)/2\Ts \}$
	and the set of side subcarriers by $\cKside =\{k \big| |\Delta f_k|> (1-\beta)/2\Ts \}$. As shown in Fig. \ref{fig:subcarrier_grouping}, there are approximately $|\cKcen| \approx \lfloor(1-\beta)K\rfloor$ central subcarriers inside $\cKcen$ and $|\cKside| \approx \lceil\beta K\rceil$ side subcarriers inside $\cKside$.  
\end{remark}
 
\vspace*{-8mm}
\subsection{Frequency domain signal model for training}
With the aforementioned system and channel model, the received signal model for training is described as follows. During the training phase, the TX will send pilot signals over several OFDM frames, so the RX can collect measurements for initial channel estimation and DL. Specifically, during the $m$-th ($m=1,2,\ldots$) training OFDM symbol, the TX will send an $\Lt\times 1$ pilot signal $\bs_m[k]$ at the $k$-th subcarrier using a frequency-flat precoder ${\bF}_m\in\bbC^{\Nt\times \Lt}$, while the RX employs a frequency-flat combiner ${\bW}_m\in\bbC^{\Nr\times \Lr}$. The received signal is then given as
\begin{align}\label{equ:yk}
\by_m[k] = \bW_m^*\bH[k] \bF_m{\bs}_m[k]  + {\bn}_m[k],
\end{align} 
where $\bn_m[k]\in \bbC^{\Lr\times 1}$ is the additive Gaussian noise vector, distributed as $\cC\cN(\bf0,\sigma^2 \breve{\bW}^*_m\breve{\bW}_m)$.
%Moreover, 
%$\bs_m[k]$ satisfies $\bbE\{\bs_m[k]\bs_m^*[k]\}=\frac{P_{\T}}{\Lt}\bI_{\Lt}$, with $P_{\T}$ the transmit power constraint. 
Generally, we can decompose the transmitted signal as $\bs_m[k]\teq\bq_m s[k]$, with $\bq_m \in\bbC^{\Lt \times 1}$ a frequency-flat training vector and $s[k]$ a frequency-dependent training symbol. In doing so, we can multiply the received signal $\by_m[k]$ by $(s[k])^{-1}$ and get a frequency-flat observation matrix at the RX as follows 
\begin{align}\label{equ:yk2}
\hspace*{-1mm}\tilde{\by}_m[k] \teq \vec\left( \by_m[k](s[k])^{-1} \right) 
 \hspace*{-1mm} = \hspace*{-1mm} (\bq_m^T\bF_m^T \kron \bW_m^*)\vec(\bH[k]) \hspace*{-1mm} + \hspace*{-1mm} \tilde{\bn}_m[k] 
\hspace*{-1mm} = \hspace*{-1mm} \bPhi_m \vec(\bH[k]) \hspace*{-1mm} + \hspace*{-1mm} \tilde{\bn}_m[k] ,
\end{align}
%\begin{align}\label{equ:yk2}
%\tilde{\by}_m[k] \teq \vec\left( \by_m[k](s[k])^{-1} \right) 
%& = (\bq_m^T\bF_m^T \kron \bW_m^*)\vec(\bH[k]) + \tilde{\bn}_m[k] \notag\\
%& = \bPhi_m \vec(\bH[k]) + \tilde{\bn}_m[k] ,
%\end{align}
where $\tilde{\bn}_m[k] \teq \vec\big({\bn}_m[k](s[k])^{-1} \big)$ and $\bPhi_m $ is defined accordingly, representing the sensing matrix for the $m$-th OFDM symbol.
To get a higher effective SNR for the received measurements, we will use training spreading to average out the noise \cite{xie2020dictionary}. Moreover, to enable the initial channel estimate, the measurements over  $M$ OFDM symbols are stacked together such that 
\begin{align}
\underbrace{\left[\tilde{\by}_1[k]^T,\ldots, \tilde{\by}_M[k]^T\right]^T}_{\tilde{\by}[k]} = \underbrace{\left[\bPhi_1^T,\ldots, \bPhi_M^T\right]^T }_{\bPhi}\bh[k] + \underbrace{\left[\tilde{\bn}_1[k]^T,\ldots, \tilde{\bn}_M[k]^T\right]^T}_{\tilde{\bn}[k]},
\end{align}
where $\tilde{\by}[k] \in \bbC^{M\Lr\times 1}$, $\bPhi\in\bbC^{M\Lr\times \Nr\Nt}$ and $\tilde{\bn}[k] \in \bbC^{M\Lr\times 1}$ are defined accordingly. Therefore, we can obtain the initial least squares (LS)  channel estimate as
%\begin{align}\label{equ:hkhat}
$\hat{\bh}[k] = \bPhi^{\dag} \tilde{\by}[k]$,
%\end{align}
while the corresponding channel matrix $\hat{\bH}[k] =\text{unvec}(\hat{\bh}[k])$. 
%\begin{align}\label{equ:hkhat}
%\hat{\bh}[k] = \bPhi^{\dag} \tilde{\by}[k],
%\end{align}
%and the corresponding channel matrix $\hat{\bH}[k] =\text{unvec}(\hat{\bh}[k])$. 

During the training phase, it becomes necessary to collect initial channel measurements at different locations across the coverage area as the RX moves around. This not only creates a large training data set for learning the hardware imperfections,  but also ensures the data set is diverse for the environment. In doing so, the learned sparsifying dictionary is not dedicated for a specific location, but adapted to the fixed hardware impairments. As  in \cite{xie2020dictionary}, this procedure of collecting measurements at multiple locations can be done at the stage of network setup, like in the case of an indoor WiFi scenario. Therefore, we assume the measurements are collected at $\Nsa$ locations and the initial channel estimate at the $u$-th location is then represented as $\hat{\bh}^{(u)}[k]$ for $u\in \cI(\Nsa)$. 
In the following, these initial channel estimates obtained at the training phases will be used for learning the hardware impairments related dictionaries.

\section{Dictionary Learning for Hardware Impairments under beam squint (DLHWBS)}
\label{sec:DLHW}
  In this section, we present the problem formulation and optimization for learning the TX and RX dictionaries adapted to hardware impairments.
Generally, the overall dictionaries are partitioned into two parts, i.e., one frequency-flat part adapting to the hardware impairments and one frequency-dependent part accounting for the beam squint effect. The learning of hardware impairment related dictionaries will take into consideration of the channel properties under beam squint at different subcarriers.
 \subsection{Formulation of the dictionary learning problem}
 As shown in \eqref{equ:Hk_two_cases}, the frequency domain channel model can be partitioned into two sets. We treat   each one separately here. 
For central subcarriers inside $\cKcen$, the channel model can be simplified so that it is more convenient and straightforward to separate the hardware impairments and beam squint impact on the channel matrix.  
Let us first rewrite the channel matrix for $k\in\cKcen$ as follows 
\begingroup
\allowdisplaybreaks
\begin{align}\label{equ:Hkin_grids}  
{\bf H}[k] 
& = \sum_{\ell=1}^L \alphal \frac{\cF(p)[\Delta f_k]}{\Ts}e^{-\j2\pi\Delta f_k \taul} \breve{\ba}_{\R,k}(\phil)\breve{\ba}^*_{\T,k}(\thetal) 
= \breve{\bA}_{\R,k}(\bm\phi) \bm\Lambda[k] \breve{\bA}_{\T,k}^*(\bm\theta)\notag\\
& \approx \breve{\bA}^{\rm v}_{\R,k}  \bm\Omega[k] \breve{\bA}_{\T,k}^{\rm v*}, 
\end{align} 
\endgroup
where $\bm\Lambda[k] \teq \diag\big\{[\alpha_1 \frac{\cF(p)[\Delta f_k]}{\Ts}e^{-\j2\pi\Delta f_k \tau_1},\ldots,\alpha_L \frac{\cF(p)[\Delta f_k]}{\Ts}e^{-\j2\pi\Delta f_k \tau_L}]^T\big\}$ is a diagonal matrix containing the frequency domain path gains, and $\breve{\bA}_{\R,k}(\bm\phi) \teq \left[ \breve{\ba}_{\R,k}(\phi_1),\ldots, \breve{\ba}_{\R,k}(\phi_L) \right] \in \bbC^{\Nr\times L} $ collects the receive array response vectors at all paths. 
Moreover, the approximation is obtained by discretizing the AoA/AoD spaces with on-grid angles, i.e., $\breve{\bA}^{\rm v}_{\R,k} \teq [\breve\ba_{\R,k}(\phi^{\rm v}_1), \ldots, \allowbreak \breve\ba_{\R,k}(\phi^{\rm v}_{\Kr})]  \in \bbC^{\Nr\times \Kr}$ and $\breve{\bA}_{\T,k}^{\rm v} \teq [\breve\ba_{\T,k}(\theta^{\rm v}_1), \ldots, \allowbreak \breve\ba_{\T,k}(\theta^{\rm v}_{\Kt})] \in \bbC^{\Nt\times \Kt}$ collect the virtual receive and transmit array response vectors evaluated on $\Kr $ quantized angles $\bm\phi^{\rm v} =\{\phi^{\rm v}_i\}_{i=1}^{\Kr}$ for AoAs and $\Kt $ quantized angles $\bm\theta^{\rm v} =\{\theta^{\rm v}_i\}_{i=1}^{\Kt}$ for AoDs, and $\bm\Omega[k] \in \bbC^{\Kr\times \Kt}$ is a sparse matrix, containing the path gains of these discrete quantized AoAs/AoDs at its non-zero elements. 
To separate the impact of hardware impairments and beam squint on channel matrix, we can rewrite the array steering vector in \eqref{equ:baRT} as 
\vspace*{4mm}
\begin{align}\label{equ:aRk}
[\check{\ba}_{\R,k}(\phi)]_{n_{\R}} & = \frac{1}{\sqrt{\Nr}}e^{-\j2\pi f_k(n_{\R}\dr+\epsilon_{\R,n_{\R}})\sin(\phi)/c}\notag\\
%& = \frac{1}{\sqrt{\Nr}}e^{-\j2\pi(f_k \epsilon_{\R,n_{\R}})\sin(\phi)/c}e^{-\j2\pi(f_k n_{\R}\dr)\sin(\phi)/c}\notag\\
& = \frac{1}{\sqrt{\Nr}}e^{-\j2\pi(f_{\rm c}\epsilon_{\R,n_{\R}}+\Delta f_k \epsilon_{\R,n_{\R}})\sin(\phi)/c}e^{-\j2\pi(f_k n_{\R}\dr)\sin(\phi)/c}\notag\\
& \approx e^{-\j2\pi(\fc\epsilon_{\R,n_{\R}})\sin(\phi)/c}\cdot \frac{1}{\sqrt{\Nr}}e^{-\j2\pi(f_k n_{\R}\dr)\sin(\phi)/c}, \\
\check{\ba}_{\R,k}(\phi) & \approx {\bf e}_{\R}(\phi)\odot  \ba_{{\rm R}, k}(\phi), 
\end{align} 
where $[{\bf e}_{\R}(\phi)]_{n_{\R}} = e^{-\j2\pi(f_{\rm c}\epsilon_{\R,n_{\R}})\sin(\phi)/c}$ comprises the hardware antenna element location imperfection effects. Therefore, the expression of $\breve{\bA}^{\rm v}_{\R,k}$ in \eqref{equ:Hkin_grids}  can be rewritten as 
\begin{align}\label{equ:ARkv} 
\breve{\bA}^{\rm v}_{\R,k} &\teq [\breve\ba_{\R,k}(\phi^{\rm v}_1), \ldots, \breve\ba_{\R,k}(\phi^{\rm v}_{\Kr})] \notag\\
& = \underbrace{\diag(\bGamma_{\R}){\bf C}_\R }_{\bD_{\R,1}} \bigg( \underbrace{ \left[ {\bf e}_{\R}(\phi^{\rm v}_1),\ldots, {\bf e}_{\R}(\phi^{\rm v}_{\Kr}) \right] }_{ \bD_{\R,2}} \odot  \underbrace{\left[\ba_{\R, k}(\phi^{\rm v}_1),\ldots,\ba_{\R, k}(\phi^{\rm v}_{\Kr}) \right]}_{\bA_{\R,k}^{\rm v} }\bigg) \notag \\
& = \bD_{\R,1}\left(\bD_{\R,2} \odot \bA_{\R,k}^{\rm v}  \right),	
\end{align}  
where the general RX dictionary $\breve{\bA}^{\rm v}_{\R,k}$ is partitioned into three parts, i.e., $\bD_{\R,1}$ includes the impacts of antenna coupling and gain/phase errors, $\bD_{\R,2}$ accounts for the antenna location perturbations, and $\bA_{\R,k}^{\rm v}\in\bbC^{\Nr\times \Kr}$ handles the beam squint effect at each subcarriers. Note that $\bD_{\R,1}$ and $\bD_{\R,2}$ are hardware impairments related dictionary components to be learned while $\bA_{\R,k}^{\rm v}$ is known for all subcarriers. Moreover, recalling the notation of antenna location error vector $\bm\epsilon_{\R}$ and the definition of $\bf e_{\R}(\phi)$ in \eqref{equ:aRk}, we can further express  $\bD_{\R,2}$ as a function of the antenna location error vector $\bm\epsilon_{\R}$, i.e., $\bD_{\R,2} = f(\bm\epsilon_\R)=  e^{-\j2\pi \fc \bm\epsilon_{\R} \cdot \sin(\bm\phi^{\rm v})^T/c   }$.
Similarly, the general TX dictionary $\breve{\bA}^{\rm v}_{\T,k}$ can be expressed as 
$\breve{\bA}^{\rm v}_{\T,k} = \bD_{\T,1}\left(\bD_{\T,2} \odot \bA_{\T,k}^{\rm v}  \right)	$ and
$\bD_{\T,2}$ is a function of the antenna location error vector $\bm\epsilon_{\T}$, i.e., $\bD_{\T,2} = f(\bm\epsilon_\T)=  e^{-\j2\pi \fc \bm\epsilon_{\T} \cdot \sin(\bm\theta^{\rm v})^T/c   }$. 
Then the vectorization of $\bH[k],k\in\cKcen$ in \eqref{equ:Hkin_grids}  can be expressed as
\begin{align} \label{equ:Hkin_grid_vec}
\bh[k] & = \vec(\bH[k]) \approx \left( \overline{\breve{\bA}_{\T,k}^{\rm v}} \otimes \breve{\bA}^{\rm v}_{\R,k}  \right) \vec(\bm\Omega[k]) \notag\\ 
& = \left( \overline{\bD_{\T,1}}\otimes \bD_{\R,1} \right)\left[ \left( \overline{\bD_{\T,2}}\otimes \bD_{\R,2} \right) \odot  \left( \overline{ \bA_{\T,k}^{\rm v} }\otimes  \bA_{\R,k}^{\rm v} \right) \right]\vec(\bm\Omega[k]).
\end{align}
 
 The channel model for side subcarriers $k\in\cKside$ in \eqref{equ:Hk_two_cases} can be represented and approximated as follows
\begin{align} \label{equ:Hkout_grids} 
\bH[k] & = \sum_{\ell=1}^L \alphal e^{-\j 2\pi\Delta f_k\taul} {\bf G}_k(\taul, \phil, \thetal) \odot( \breve{\ba}_{\R,k}(\phil) \breve{\ba}^*_{\T,k}(\thetal)) \nn\\
%& \approx  \sum_{\ell=1}^L \sum_{i=1}^{\Kr}\sum_{j=1}^{\Kt}b_{\ell,i,j} e^{-\j 2\pi f_k\taul} {\bf G}_k(\taul, \phi_{i}^{\rm v}, \theta_{j}^{\rm v}) \odot( \breve{\ba}_{\R,k}(\phi_{i}^{\rm v}) \breve{\ba}^*_{\T,k}(\theta_{j}^{\rm v})) \nn\\
& \approx  \sum_{l=1}^{L_{\rm v}} \sum_{i=1}^{\Kr}\sum_{j=1}^{\Kt}b_{l,i,j} e^{-\j 2\pi\Delta f_k\tau_l^{\rm v}} {\bf G}_k(\tau_l^{\rm v}, \phi_{i}^{\rm v}, \theta_{j}^{\rm v}) \odot( \breve{\ba}_{\R,k}(\phi_{i}^{\rm v}) \breve{\ba}^*_{\T,k}(\theta_{j}^{\rm v}))
\end{align} 
where the approximation is obtained by first discretizing the AoA/AoD spaces as did in \eqref{equ:Hkin_grids}  and then discretizing the delay space with $L_{\rm v}$ on-grid delays $\{\tau_l^{\rm v}\}_{l=1}^{L_{\rm v}}$. Moreover, $b_{l,i,j}$ is the corresponding path gain on each pair of discretized AoA/AoD/delay grids, which theoretically is nonzero only at the discrete AoA/AoD/delay bin $(\phi^{\text{v}}_i, \theta^{\text{v}}_j,\tau_l^{\rm v})$ corresponding to the true AoA/AoD/delay $(\phil,\thetal,\taul)$. Then the vectorization of $\bH[k],k\in\cKside$ can be given as
\begin{align} \label{equ:Hkout_grids_vec} 
	 \bh[k] & = \vec(\bH[k]) = \sum_{\ell=1}^L \alphal e^{-\j 2\pi\Delta f_k\taul} \vec({\bf G}_k(\taul, \phil, \thetal)) \odot(\overline{\breve{\ba}_{\T,k}(\thetal)}\kron \breve{\ba}_{\R,k}(\phil) ) \nn\\
	 & \approx \sum_{l=1}^{L_{\rm v}} \Big[{\bm\Psi}_{k}(\tau_l^{\rm v}, \bm\phi^{\text{v}}, \bm\theta^{\text{v}})  \odot \left( \overline{\breve{\bA}_{\T,k}^{\rm v}} \otimes \breve{\bA}^{\rm v}_{\R,k}  \right)\Big] \bm b_{l}, 
\end{align} 
where ${\bm\Psi}_{k}(\tau_l^{\rm v}, \bm\phi^{\text{v}}, \bm\theta^{\text{v}}) \in \bbC^{\Nr\Nt \times \Kr\Kt}$ and its columns are defined as 
\begin{align} 
	\big[{\bm\Psi}_{k}(\tau_l^{\rm v}, \bm\phi^{\text{v}}, \bm\theta^{\text{v}})\big]_{:,(j-1)\Kt+i} = 
	e^{-\j 2\pi\Delta f_k\tau_l^{\rm v}} \vec({\bf G}_k(\tau_l^{\rm v}, \phi^{\text{v}}_i, \theta^{\text{v}}_j)), \ \forall i=1,\ldots,\Kr,j=1,\ldots,\Kt,
\end{align} 
and $\bm b_{l}\in\bbC^{\Kr\Kt\times 1}$ collects $b_{l,i,j}$, satisfying $\|\bm b_{l}\|_0\leq 1$ and $\sum_{l=1}^{L_{\rm v}}\|\bm b_{l}\|_0 \leq L$.

\begin{remark} 
	Comparing the channel approximation expressions for central subcarriers in \eqref{equ:Hkin_grids} and for side subcarriers in \eqref{equ:Hkout_grids}, it is clear that the additional distortion matrix $\bG_k(\taul,\phil,\thetal)$ for side subcarriers entangles the three parameters of delay, AoA and AoD, such that a combination of discretized delay, AoA and AoD grids is needed. This will increase the sparsifying dictionary dimension significantly and induce overwhelming computational complexity during the CS recovery of these parameters. Instead, for the central subcarriers, the discretization of delay, AoA and AoD is decoupled without the additional distortion matrix and thus even one-dimensional search can be done sequentially for each parameter, which will help reduce the computational complexity to a large extent. 
%	Motivated by these, we will propose a new sparse coding algorithm for parameter recovery in the next section. 
\end{remark}
	
Next, as per the approximate channel models in \eqref{equ:Hkin_grids}-\eqref{equ:Hkin_grid_vec} as well as \eqref{equ:Hkout_grids}-\eqref{equ:Hkout_grids_vec}, we can formulate the final DL problem for hardware impairments.
%For ease of expression, we denote the unknown channel parameter set for the $l$-th path of the $u$-th channel measurement by $\bm\varphi_{l}^{(u)} \teq\{\alpha_{l}^{(u)},\phi_{l}^{(u)},\theta_{l}^{(u)},\tau_{l}^{(u)}\}$ for $l=1,\ldots,L^{(u)}, \forall u\in\cI(\Nsa)$. 
Specifically, stacking the initial channel estimates at all subcarriers from all locations $u\in\cI(\Nsa)$, the problem formulation of DLHW can be expressed as
\begingroup
\allowdisplaybreaks
\begin{align} \label{equ:DL2} 
& \min_{\substack{\bD_{\T,1},\bD_{\T,2},\bD_{\R,1},\bD_{\R,2},\\ \bm\Omega^{(u)}[k], \bm b^{(u)}_{l} } }  
\ \sum_{u\in\cI(\Nsa)}\sum_{k\in\cKcen} \left\|\hat{\bh}^{(u)}[k] -  \left( \overline{\breve{\bA}_{\T,k}^{\rm v}} \otimes \breve{\bA}^{\rm v}_{\R,k}  \right)\vec(\bm\Omega^{(u)}[k]) \right\|_F^2  \nn\\ 
&\kern 75pt + \sum_{u\in\cI(\Nsa)}\sum_{k\in \cKside } \Big\| \hat{\bh}^{(u)}[k] - \sum_{l=1}^{L_{\rm v}} \Big[{\bm\Psi}_{k}(\tau_{l}^{\rm v}, \bm\phi^{\text{v}}, \bm\theta^{\text{v}})  \odot \left( \overline{\breve{\bA}_{\T,k}^{\rm v}} \otimes \breve{\bA}^{\rm v}_{\R,k}  \right)\Big] \bm b_{l}^{(u)}\Big\|^2_F, \nn \\
& \kern 20pt \text{subject to}\quad\quad  \big\|\vec(\bm\Omega^{(u)}[k])\big\|_{0}\leq L^{(u)}; \quad \big\|\bm b_{l}^{(u)}\big\|_{0}\leq 1, \quad \sum_{l=1}^{L_{\rm v}} \big\|\bm b_{l}^{(u)}\big\|_{0}\leq L^{(u)}. 
\end{align} 
\endgroup
Note that the two sum terms in \eqref{equ:DL2} are format-consistent as $\big[{\bm\Psi}_{k}(\tau_{l}^{\rm v}, \bm\phi^{\text{v}},  \bm\theta^{\text{v}})\big]_{:,(j-1)\Kt+i} = 
\frac{\cF(p)[f_k]}{\Ts}e^{-\j 2\pi f_k\tau_{l}^{\rm v}}\bm 1_{\Nr\Nt}$ for central subcarriers $k\in\cKcen$. We will see that the simplified expression for central subcarriers can help facilitate the derivations of DL algorithms.  

\subsection{Speeding up the dictionary learning with a new sparse coding algorithm}
The optimization problem in \eqref{equ:DL2} is not jointly convex with respect to the variables $\bD_{\T,1}$, $\bD_{\T,2}$, $\bD_{\R,1}$, $\bD_{\R,2}$, $\bm\Omega^{(u)}[k]$, and $ \bm b^{(u)}_{l}$, but it can still be solved by the alternating optimization techniques. As in the typical DL problems \cite{KSVD}, the optimization of \eqref{equ:DL2} is split into two stages: sparse coding and dictionary update.
\subsubsection{Sparse coding stage} \label{sec:sparsecoding}
In this stage, we fix all dictionary parts and update the channel coefficients $\bm\Omega^{(u)}[k], \forall k\in\cKcen, u\in\cI(\Nsa)$ and $\bm b_{l}^{(u)}, \forall k\in\cKside, u\in\cI(\Nsa)$.
% as well as the path parameters  $\bm\varphi_{\ell}^{(u)},\ell=1,\ldots,L^{(u)},u\in\cI(\Nsa)$. 
Specifically, for each $u\in\cI(\Nsa)$, the optimization problem of \eqref{equ:DL2} is reduced to (omitting the superscript $^{(u)}$ for simplicity)
\begingroup
\allowdisplaybreaks 
\begin{align} \label{equ:sparsecoding2}
& \min_{ \bm\Omega[k], \bm b_{l} }  
\quad  \sum_{k\in\cKcen} \left\|\hat{\bh}[k] -  \left( \overline{\breve{\bA}_{\T,k}^{\rm v}} \otimes \breve{\bA}^{\rm v}_{\R,k}  \right)\vec(\bm\Omega[k]) \right\|_F^2  \nn\\ 
&\kern 29pt +  \sum_{k\in \cKside } \Big\| \hat{\bh}[k] - \sum_{l=1}^{L_{\rm v}} \Big[{\bm\Psi}_{k}(\tau_{l}^{\rm v}, \bm\phi^{\text{v}}, \bm\theta^{\text{v}})  \odot \left( \overline{\breve{\bA}_{\T,k}^{\rm v}} \otimes \breve{\bA}^{\rm v}_{\R,k}  \right)\Big] \bm b_{l}\Big\|^2_F, \nn \\
& \text{subject to}\quad\quad  \big\|\vec(\bm\Omega[k])\big\|_{0}\leq L; \quad \big\|\bm b_{l}\big\|_{0}\leq 1, \quad \sum_{l=1}^{L_{\rm v}} \big\|\bm b_{l}\big\|_{0}\leq L.  
\end{align}   
\endgroup  
Generally, this problem can be optimized by various CS techniques, such as orthogonal matching pursuit (OMP) \cite{OMP}, simultaneous OMP (SOMP) \cite{SOMP}, or simultaneous weighted OMP (SWOMP) \cite{rodriguez2018frequency}, to name a few. 
In this paper,  we propose a new sparse coding algorithm to solve \eqref{equ:sparsecoding2}, named \textit{Dictionary Adaptive OMP under beam squint (DA-OMP-BS)} as shown in Algorithm \ref{alg:DA-OMP-BS}. This algorithm exploits the following two important properties of the channel models under beam squint, which will enable low-complexity and high-accuracy recovery of the channel coefficients and parameters:
\begin{itemize}
	\item First, the common sparsity support property of channel vectors across subcarriers will still be considered but with frequency-dependent dictionaries accounting for beam squint impacts. 
	For SOMP and OMP in previous work, the common sparsity support of channel coefficients between subcarriers is assumed under the same dictionary for all subcarriers. This is an approximate result for the case without considering beam squint effect. When the steering vector at the center frequency is used for all subcarriers, there exist approximation errors no matter how significant the beam squint effect is. 
%	Therefore, the previous assumption of common sparsity support is considered as valid when beam squint impact is negligible. 
	Under beam squint circumstances, we can exploit the true property of common sparsity support between subcarriers but with different frequency-dependent dictionaries accounting for the beam squint impacts. The principle behind this argument is that the physical AoAs and AoDs associated with propagation paths are constant and independent of subcarriers, and thereby when the channel vector at each subcarrier is projected to the corresponding sparsifying dictionary, only a few bins out of the $\Kr$ or $\Kt$ virtual angular bins corresponding to the physical AoAs and AoDs are nonzero. 
	%This is exactly the true property of common sparsity support between subcarriers under beam squint, and thus the dictionaries should be adapted at different subcarriers accounting for the beam squint impact.
	\item Second, as mentioned above, the additional distortion matrix at side subcarriers tangles the three parameters of delay, AoA and AoD. Then if traditional SOMP or OMP is directly applied for \eqref{equ:sparsecoding2}, the overall sparsifying dictionary will be a three-dimensional (3D) dictionary \cite{VenAlkPre:Channel-Estimation-for-Hybrid:17} and the dimension of this 3D dictionary $\allowbreak [{\bm\Psi}_{k}(\tau_{1}^{\rm v}, \bm\phi^{\text{v}}, \bm\theta^{\text{v}})  \odot \big( \overline{\breve{\bA}_{\T,k}^{\rm v}} \otimes \breve{\bA}^{\rm v}_{\R,k}  \big),\ldots, \\{\bm\Psi}_{k}(\tau_{L_{\rm v}}^{\rm v}, \bm\phi^{\text{v}}, \bm\theta^{\text{v}})  \odot \big( \overline{\breve{\bA}_{\T,k}^{\rm v}} \otimes \breve{\bA}^{\rm v}_{\R,k}  \big)]$ will be $\Nr\Nt \times L_{\rm v}\Kr\Kt$, proportional to the product of the numbers of delay/AoA/AoD grids. This will  induce overwhelming complexity. As for the newly proposed DA-OMP-BS, it will first exploit the channel models at central subcarriers to obtain initial estimates of the delay/AoA/AoD parameters. Without the impact of additional distortion matrix at central subcarriers, this can be done even by one-dimensional search over each parameter space iteratively. These initial estimates of parameters are then used to reduce the effective dictionary dimension at side subcarriers and compensate the distortions induced by beam squint, 
	so that all the subcarriers can be collected to improve the estimates of delays/AoAs/AoDs again. In doing so, the sparse coding of \eqref{equ:sparsecoding2} can be solved with much lower complexity and higher accuracy. 
\end{itemize} 
In line of these ideas, the DA-OMP-BS is expected to outperform SOMP and OMP without considering beam squint. We summarize the procedure of DA-OMP-BS in Algorithm \ref{alg:DA-OMP-BS}. 
Once the sparse coding stage of \eqref{equ:sparsecoding2} is done, the coefficients $\bm\Omega^{(u)}[k], k\in\cKcen$ and $\bm b_{l}^{(u)},  k\in\cKside$ as well as the estimates of path gains, delays, AoAs, and AoDs  $\hat{\bm\varphi}_{\hat{l}}^{(u)} \teq\{\hat{\alpha}_{\hat{l}}^{(u)},\hat{\phi}_{\hat{l}}^{(u)},\hat{\theta}_{\hat{l}}^{(u)},\hat{\tau}_{\hat{l}}^{(u)}\}$ for $\hat{l}=1,\ldots,\hat{L}^{(u)}, \forall u\in\cI(\Nsa)$ can be obtained, which will be used to update the hardware impairments related dictionaries in the next subsections.

\begin{algorithm}[!ht]
	\caption{: DA-OMP-BS algorithm for sparse coding}
	\label{alg:DA-OMP-BS} 
	\begin{algorithmic}[1]
		\State{\textbf{procedure} DA-OMP-BS ($  {\by}^{(u)}[k]$, $\bPhi$, $\bD_{\T,1}$, $\bD_{\T,2}$, $\bD_{\R,1}$, $\bD_{\R,2}, \Kr,\Kt,L_{\rm v}$)}
		\State \quad \textbf{Construct $\breve{\bA}_{\R,k}^{\rm v}$ and $\breve{\bA}_{\T,k}^{\rm v}$ at each subcarrier with input dictionaries}
		\State \quad \textbf{Compute the whitened equivalent sensing matrix and received signals} \\
		\qquad $ \bL_{\rm w} = \text{blkdiag}\left\{(\breve\bW_1^*\breve\bW_1)^{-1/2},\ldots, (\breve\bW_M^*\breve\bW_M)^{-1/2} \right\}$ \quad and 
		\quad $\bPhi_{\rm w} = \bL_{\rm w}\bPhi$ \\
		%		\qquad \quad $\bUpsilon_{\text{w}}[k] = \bL_{\rm w}\bPhi(\overline{\breve{\bA}_{\T,k}^{\text{v}}}\kron \breve{\bA}_{\R,k}^{\text{v}}), \ \forall k\in\cKcen$\\
		\qquad $\by_{\text{w}}^{(u)}[k] = \bL_{\rm w} \by^{(u)}[k], \ \text{for}\ k=0,\ldots,K-1,\ u\in\cI(\Nsa)$  
		\State \quad \textbf{for} $u = 0,\ldots,\Nsa - 1$ \textbf{do}	
		\State \qquad \textbf{Initialize the residual vectors and set of estimated parameters} \\		  
		\qquad \quad $\br[k] = \bPhi_{\rm w}^*\by_{\text{w}}^{(u)}[k], \ \forall k=0,\ldots,K-1 $ and $\hat{L}^{(u)}=0,\ \hat{\bm\varphi}^{(u)}_{l}=\emptyset$ 
		\State \qquad \textbf{while} $\text{MSE}>\epsilon$ \textbf{do:} \\
		\qquad \quad \text{Define:} $O(\tau, \phi, \theta, \mathcal{K}_{\rm set}) = \frac{ \big\|\sum_{k\in\mathcal{K}_{\rm set}} ( \overline{\breve{\ba}_{\T,k}^{\rm v}(\theta)}\kron \breve{\ba}^{\rm v}_{\R,k}(\phi))^* (\br[k]e^{\j2\pi\Delta f_k \tau} )\big\|_F^2 }{ \big\|\sum_{k\in\mathcal{K}_{\rm set}} ( \overline{\breve{\ba}_{\T,k}^{\rm v}(\theta)}\kron \breve{\ba}^{\rm v}_{\R,k}(\phi))^* ( \bPhi_{\rm w}^* e^{\j2\pi\Delta f_k \tau} )\big\|_F^2}$\\
		\qquad \quad \textbf{Initial parameter estimation using central subcarriers}\\ 
		\qquad \qquad \text{delay:} $\hat{\tau} =\argmax_{\tau_l^{\rm v}}\ \big\|\sum_{k\in\cKcen}\br[k]e^{\j2\pi\Delta f_k\tau_l^{\rm v}}\big\|_F^2$\\
		\qquad \qquad \text{AoA:} $\hat{\phi} =\argmax_{\phi_i^{\rm v}}\ \frac{ \big\|\sum_{k\in\cKcen} (\bI_{\Nt}\kron \breve{\ba}^{\rm v}_{\R,k}(\phi_i^{\rm v}))^* (\br[k]e^{\j2\pi\Delta f_k \hat{\tau}} )\big\|_F^2 }{ \big\|\sum_{k\in\cKcen} (\bI_{\Nt}\kron \breve{\ba}^{\rm v}_{\R,k}(\phi_i^{\rm v}))^* (\bPhi_{\rm w}^*e^{\j2\pi\Delta f_k \hat{\tau}} )\big\|_F^2 
		}$\\
		\qquad \qquad \text{AoD:} $\hat{\theta} = \argmax_{\theta_j^{\rm v}}\ O(\hat{\tau}, \hat{\phi}, \theta_j^{\rm v}, \mathcal{K}_{\rm cen})$\\
		\qquad \quad\textbf{Refine the initial parameter estimates}\\
		\qquad \qquad \text{delay:} $\hat{\tau} = \argmax_{\tau_l^{\rm v}}\ O(\tau_l^{\rm v}, \hat{\phi}, \hat{\theta}, \mathcal{K}_{\rm cen})$\\
		\qquad \qquad \text{AoA:} $\hat{\phi} = \argmax_{\phi_i^{\rm v}}\ O(\hat{\tau}, \phi_i^{\rm v}, \hat{\theta}, \mathcal{K}_{\rm cen})$\\
		\qquad \qquad \text{AoD:} $\hat{\theta} = \argmax_{\theta_j^{\rm v}}\ O(\hat{\tau}, \hat{\phi}, \theta_j^{\rm v}, \mathcal{K}_{\rm cen})$\\		
		\qquad \quad\textbf{Distortion compensation at side subcarriers }\\
		\qquad \qquad $\br[k] = \diag\big\{\vec(\bG_k(\hat{\tau},\hat{\phi},\hat{\theta}))\big\}^{-1} \br[k],\ \forall  k \in \cKside$\\
		\qquad \quad\textbf{Update parameter estimates using all subcarriers}\\
		\qquad \qquad \text{delay:} $\hat{\tau} = \argmax_{\tau_l^{\rm v}}\ O(\tau_l^{\rm v}, \hat{\phi}, \hat{\theta}, \mathcal{K}_{\rm cen}\cup\mathcal{K}_{\rm side})$\\
		\qquad \qquad \text{AoA:} $\hat{\phi} = \argmax_{\phi_i^{\rm v}}\ O(\hat{\tau}, \phi_i^{\rm v}, \hat{\theta}, \mathcal{K}_{\rm cen}\cup\mathcal{K}_{\rm side})$\\
		\qquad \qquad \text{AoD:} $\hat{\theta} = \argmax_{\theta_j^{\rm v}}\ O(\hat{\tau}, \hat{\phi}, \theta_j^{\rm v}, \mathcal{K}_{\rm cen}\cup\mathcal{K}_{\rm side})$\\			
		\qquad \quad\textbf{Update the set of estimated parameters}: 
		\ $\hat{L}^{(u)}= \hat{L}^{(u)} + 1,\ \hat{\bm\varphi}^{(u)}_{l}=\{\hat{\tau},\hat{\phi},\hat{\theta}\}$	
		\algstore{myalg}
		%%		\State \qquad \quad \textbf{end while} 		 
		%		\State \quad \textbf{end for} 
		%		\State \quad \textbf{Output:} The coefficients   $\vec(\hat{\bm\Omega}^{(u)}[k]), \ \forall k\in\cKcen, u\in\cI(\Nsa)$	
	\end{algorithmic}
\end{algorithm} 

\begin{algorithm}[!ht] 
	\ContinuedFloat
	\caption{: DA-OMP-BS algorithm for sparse coding (continued)}  
	\begin{algorithmic}[1]
		\algrestore{myalg}  
		\State \qquad \quad \textbf{Update the path gains $\hat{\alpha}_l$ and coefficient vectors by minimizing} \\
		 \qquad \qquad $\sum_{k=0}^{K-1}\big\|\by_{\text{w}}^{(u)}[k]{-}\bPhi_{\rm w}\sum_{l=1}^{\hat{L}^{(u)} } \alpha_l e^{-\j 2\pi\Delta f_k\hat{\tau}_l} \vec({\bf G}_k(\hat{\tau}_l, \hat{\phi}_l, \hat{\theta}_l)) \odot(\overline{\breve{\ba}_{\T,k}(\hat{\theta}_l)}\kron \breve{\ba}_{\R,k}(\hat{\phi}_l) ) \big\|_2^2$ \\
		 \qquad \quad \textbf{Update the residual for each subcarrier} \\
		 \qquad \qquad $\br[k] {= }\bPhi_{\rm w}^*\big(\by_{\text{w}}^{(u)}[k]{-}\bPhi_{\rm w}\sum_{l=1}^{\hat{L}^{(u)} } \hat{\alpha}_l e^{-\j 2\pi\Delta f_k\hat{\tau}_l} \vec({\bf G}_k(\hat{\tau}_l, \hat{\phi}_l, \hat{\theta}_l)) \odot(\overline{\breve{\ba}_{\T,k}(\hat{\theta}_l)}\kron \breve{\ba}_{\R,k}(\hat{\phi}_l) ) \big)$ \\
		 \qquad \quad \textbf{Update the current MSE:}  $ \text{MSE} {=}\frac{1}{M\Lr K} \sum_{k=0}^{K-1} \|(\bPhi^*_{\rm w})^\dag\br[k]\|_2^2 $	 
		\State \qquad \textbf{end while}   		 
		\State \quad \textbf{end for} 
		\State \quad \textbf{Output:} $\hat{\bm\varphi}^{(u)}_{l}{=}\{\hat{\alpha}_l,\hat{\tau}_l,\hat{\phi}_l,\hat{\theta}_l\}$, $\vec(\hat{\bm\Omega}^{(u)}[k])$ and $ \bm b_{l}^{(u)}$, for $u\in\cI(\Nsa)$ 	
	\end{algorithmic} 
\end{algorithm} 
 
\subsubsection{Dictionary update stage} Next, we fix the channel coefficients and path parameters in \eqref{equ:DL2} in preparation of updating the hardware impairments related dictionaries $\bD_{\T,1},\bD_{\T,2},\bD_{\R,1},\bD_{\R,2}$. Note that due to the special structure in this problem that the TX and RX dictionaries are entangled with the beam squint effect at different subcarriers, the typical dictionary update algorithms, like the method of optimal directions (MOD) \cite{MOD} or K-SVD \cite{KSVD}, cannot be directly used.  
Therefore, we apply alternating optimization in this sub-stage as well to subsequently update the four dictionaries. 

For the update of $\bD_{\R,1}$, the problem of \eqref{equ:DL2} can be reduced to
\begingroup 
\allowdisplaybreaks
\begin{align} 
&\min_{ \bD_{\R,1}} \quad \sum_{u\in\cI(\Nsa)}\sum_{k\in\cKcen} \bigg\|\hat{\bH}^{(u)}[k] -  
\bD_{\R,1} \underbrace{\left(\bD_{\R,2} \odot \bA_{\R,k}^{\rm v}  \right) \bm\Omega^{(u)}[k]  \left(\bD_{\T,2} \odot \bA_{\T,k}^{\rm v}  \right)^*\bD_{\T,1}^*}_{\bX_{\R,1}^{(u)}[k]} \bigg\|^2_F \notag \\
& + \sum_{u\in\cI(\Nsa)}\sum_{k\in\cKside}\bigg\|
		\hat{\bH}^{(u)}[k] - \sum_{\hat{l}=1}^{\hat{L}^{(u)}} \hat{\alpha}^{(u)}_{\hat{l}} e^{-\j 2\pi\Delta f_k\hat{\tau}^{(u)}_{\hat{l}}} {\bf G}_k(\hat{\tau}^{(u)}_{\hat{l}}, \hat{\phi}^{(u)}_{\hat{l}}, \hat{\theta}^{(u)}_{\hat{l}})\odot \big(\breve{\ba}_{\R,k}(\hat{\phi}^{(u)}_{\hat{l}}) \breve{\ba}_{\T,k}^{*}(\hat{\theta}^{(u)}_{\hat{l}}) \big) \bigg\|_F^2.
\end{align}
\endgroup
which is equivalent to 
\begingroup 
\allowdisplaybreaks
\begin{align}\label{equ:DR1_obj2}
&\min_{ \bD_{\R,1}} \quad \sum_{u\in\cI(\Nsa)}\sum_{k\in\cKcen} \bigg\|\hat{\bH}^{(u)}[k] -  
\bD_{\R,1} \underbrace{\left(\bD_{\R,2} \odot \bA_{\R,k}^{\rm v}  \right) \bm\Omega^{(u)}[k]  \left(\bD_{\T,2} \odot \bA_{\T,k}^{\rm v}  \right)^*\bD_{\T,1}^*}_{\bX_{\R,1}^{(u)}[k]} \bigg\|^2_F \notag \\
& \kern 20pt + \sum_{u\in\cI(\Nsa)}\sum_{k\in\cKside}\bigg\|
\hat{\bH}^{(u)}[k] - \sum_{\hat{l}=1}^{\hat{L}^{(u)}} \hat{\alpha}^{(u)}_{\hat{l}} e^{-\j 2\pi\Delta f_k\hat{\tau}^{(u)}_{\hat{l}}}  {\bf G}_k(\hat{\tau}^{(u)}_{\hat{l}}, \hat{\phi}^{(u)}_{\hat{l}}, \hat{\theta}^{(u)}_{\hat{l}})  \notag\\
& \kern 100pt \odot \Big(\bD_{\R,1} \underbrace{\big(\bee_{\R}(\hat{\phi}^{(u)}_{\hat{l}})\odot   {\ba}_{\R,k}(\hat{\phi}^{(u)}_{\hat{l}}) \big)
	\big(\bee_{\T}(\hat{\theta}^{(u)}_{\hat{l}})\odot   {\ba}_{\T,k}(\hat{\theta}^{(u)}_{\hat{l}}) \big)^* \bD_{\T,1}^* }_{\bX_{\R,1,\hat{l}}^{(u)}[k]}  \Big) \bigg\|_F^2,
\end{align}
\endgroup
where $\bX_{\R,1}^{(u)}[k]$ and $\bX_{\R,1,\hat{l}}^{(u)}[k]$ are defined accordingly for ease of expression.
To update $\bD_{\R,1}$, we need to calculate the derivative of the objective function with respect to $\bD_{\R,1}$, i.e., $\displaystyle \frac{\partial J}{\partial \bD_{\R,1}}$, which is expressed as follows (A proof is provided in Appendix II.)
\begingroup
\allowdisplaybreaks
 \begin{align} \label{equ:derivative_DR1_final}
& \frac{\partial J }{\partial \bD_{\R,1}}  =   - \sum_{u\in\cI(\Nsa)}\sum_{k\in\cKcen} \overline{( \hat{\bH}^{(u)}[k]- \bD_{\R,1}\bX_{\R,1}^{(u)}[k]  )}(\bX_{\R,1}^{(u)}[k])^T 
  - \sum_{u\in\cI(\Nsa)}\sum_{k\in\cKside} \sum_{\hat{l}=1}^{\hat{L}^{(u)}} \hat{\alpha}^{(u)}_{\hat{l}} e^{-\j 2\pi\Delta f_k\hat{\tau}^{(u)}_{\hat{l}}} \notag\\
 & \kern 40pt \cdot \bigg[\Big(\overline{\hat{\bH}^{(u)}[k]} -  \sum_{\hat{l}=1}^{\hat{L}^{(u)}} \overline{\hat{\alpha}^{(u)}_{\hat{l}} e^{-\j 2\pi\Delta f_k\hat{\tau}^{(u)}_{\hat{l}}}  {\bf G}_k(\hat{\tau}^{(u)}_{\hat{l}}, \hat{\phi}^{(u)}_{\hat{l}}, \hat{\theta}^{(u)}_{\hat{l}})\odot \bD_{\R,1}\bX_{\R,1,\hat{l}}^{(u)}[k]} \Big) \notag\\
 & \kern 40 pt\odot  {\bf G}_k(\hat{\tau}^{(u)}_{\hat{l}}, \hat{\phi}^{(u)}_{\hat{l}}, \hat{\theta}^{(u)}_{\hat{l}})\bigg]  \cdot \big(  \bX_{\R,1,\hat{l}}^{(u)}[k] \big)^T.
 \end{align}
 \endgroup
Therefore, the update of  $\bD_{\R,1}$ can be obtained by (stochastic) gradient decent as 
\begin{align}\label{equ:DR1_update1}
\left[ \bD_{\R,1}\right]^{\rm new} = \left[ \bD_{\R,1}\right]^{\rm old} - \eta \frac{\partial J}{\partial  {\bD_{\R,1}} },
\end{align}
where $\eta$ is the step-size of gradient descent and can be determined by backtracking line search.
Similarly, the update of $\bD_{\T,1}$ can be obtained.
  
For the update of $\bD_{\R,2} = f(\bm\epsilon_{\R})\teq e^{-\j2\pi \fc \bm\epsilon_{\R} \cdot \sin(\bm\phi^{\rm v})^T/c   }$, it is equivalent to updating  $\bm\epsilon_{\R}$. By stacking all subcarriers, we have the objective of updating $\bm\epsilon_{\R}$  as follows
\begingroup
\allowdisplaybreaks
\begin{align}\label{equ:DR2_obj2} 
&\min_{ \bm\epsilon_{\R}} \sum_{u\in\cI(\Nsa)} \sum_{k\in\cKcen} \bigg\|\underbrace{\bD_{\R,1}^\dag \hat{\bH}^{(u)}[k]\bD_{\T,1}^{*\dag} }_{\bY_{\R,2}^{(u)}[k] } -  
\left(\bD_{\R,2} \odot \bA_{\R,k}^{\rm v}  \right) \underbrace{ \bm\Omega^{(u)}[k]  \left(\bD_{\T,2} \odot \bA_{\T,k}^{\rm v}  \right)^*}_{\bX_{\R,2}^{(u)}[k]} \bigg\|^2_F\notag\\
& \kern 10pt + \sum_{u\in\cI(\Nsa)}\sum_{k\in\cKside}\bigg\|
\hat{\bH}^{(u)}[k] - \sum_{\hat{l}=1}^{\hat{L}^{(u)}} \hat{\alpha}^{(u)}_{\hat{l}} e^{-\j 2\pi\Delta f_k\hat{\tau}^{(u)}_{\hat{l}}}  {\bf G}_k(\hat{\tau}^{(u)}_{\hat{l}}, \hat{\phi}^{(u)}_{\hat{l}}, \hat{\theta}^{(u)}_{\hat{l}})  \notag\\
& \kern 70pt \odot \Big(\bD_{\R,1} \big(\bee_{\R}(\hat{\phi}^{(u)}_{\hat{l}})\odot   {\ba}_{\R,k}(\hat{\phi}^{(u)}_{\hat{l}}) \big)
	\underbrace{\big(\bee_{\T}(\hat{\theta}^{(u)}_{\hat{l}})\odot   {\ba}_{\T,k}(\hat{\theta}^{(u)}_{\hat{l}}) \big)^* \bD_{\T,1}^* }_{\bX_{\R,2,\hat{l}}^{(u)}[k]}  \Big) \bigg\|_F^2,
\end{align} 
\endgroup
where $\bX_{\R,2}^{(u)}[k]$ and $\bX_{\R,2,\hat{l}}^{(u)}[k]$ are defined accordingly for ease of expression. 
To update $\bm\epsilon_{\R}$, we need to calculate the derivative of the objective function with respect to $\bm\epsilon_{\R}$, which can be expressed as (A proof is given in Appendix III)
\begingroup
\allowdisplaybreaks
\begin{align} \label{equ:epsilon_final}
& \frac{\partial J}{\partial \bm\epsilon_\R} =  \sum_{u\in\cI(\Nsa)}\sum_{k\in\cKcen} 2\cR\left\{ 
\left\{ \frac{\partial J_k^{(u)}}{\partial  \bD_{\R,2} } \odot \bD_{\R,2} \right\}\cdot \frac{-\j2\pi \fc \sin(\bm\phi^{\rm v})}{c}\right\} \notag\\
& - \sum_{u\in\cI(\Nsa)}\sum_{k\in\cKside}  2\cR\Bigg\{\sum_{\hat{l}=1}^{\hat{L}^{(u)}} \hat{\alpha}^{(u)}_{\hat{l}} e^{-\j 2\pi\Delta f_k\hat{\tau}^{(u)}_{\hat{l}}} \cdot \frac{-\j2\pi\fc \sin(\hat{\phi}^{(u)}_{\hat{l}})}{c} \cdot \bigg[ \bD_{\R,1}^T  \notag\\
& \cdot
\bigg(\Big(\overline{\hat{\bH}^{(u)}[k]} - \sum_{\hat{l}=1}^{\hat{L}^{(u)}} \overline{\hat{\alpha}^{(u)}_{\hat{l}} e^{-\j 2\pi\Delta f_k\hat{\tau}^{(u)}_{\hat{l}}}  {\bf G}_k(\hat{\tau}^{(u)}_{\hat{l}}, \hat{\phi}^{(u)}_{\hat{l}}, \hat{\theta}^{(u)}_{\hat{l}})\odot \big(\bD_{\R,1} \big(\bee_{\R}(\hat{\phi}^{(u)}_{\hat{l}})\odot   {\ba}_{\R,k}(\hat{\phi}^{(u)}_{\hat{l}}) \big) \bX_{\R,2,\hat{l}}^{(u)}[k] \big)} \Big) \notag\\
& \odot {\bf G}_k(\hat{\tau}^{(u)}_{\hat{l}}, \hat{\phi}^{(u)}_{\hat{l}}, \hat{\theta}^{(u)}_{\hat{l}}) \bigg) (\bX_{\R,2,\hat{l}}^{(u)}[k])^T \bigg] \odot \big(\bee_{\R}(\hat{\phi}^{(u)}_{\hat{l}}) \odot {\ba}_{\R,k}(\hat{\phi}^{(u)}_{\hat{l}})\big) \Bigg\}.
\end{align}
\endgroup
Therefore, the update of $\bD_{\R,2}$ is given as 
\begin{align}\label{equ:DR2_update2}
\left[ \bD_{\R,2}\right]^{\rm new} = f([\bm\epsilon_{\R}]^{\rm new}) =  f\left([\bm\epsilon_{\R}]^{\rm old} -\eta \frac{\partial J}{\partial \bm\epsilon_\R}\right).
\end{align}
Similarly, the update of $\bD_{\T,2}$ can be obtained.  
 
\begin{algorithm}[t]
	\caption{: Dictionary learning for hardware impairments under beam squint (DLHWBS)} 
	\label{alg:DLHW} 
	\begin{itemize} 
		\item \textbf{Input:} Initial channel estimates $\hat{\bh}^{(u)}[k],\forall k=0,\ldots,K-1, u\in\cI(\Nsa) $. 
		\item \textbf{Initialization:} Set the dictionary matrices $\bD_{\R,1}\in\bbC^{\Nr \times \Nr}$ and $\bD_{\T,1}\in\bbC^{\Nt \times \Nt}$ using measurement data based on DIA algorithm \cite{RusDum:An-initialization-strategy-for-the-dictionary:14}, and set 
		the dictionary matrices $\bD_{\T,2} $ and $\bD_{\R,2} $ as all-one matrices.
		\item \textbf{While} \textit{not converge} \textbf{do}
		\item[] \quad \quad 1. \textit{Sparse coding stage}: Fixing all dictionaries, solve \eqref{equ:sparsecoding2} using Algorithm \ref{alg:DA-OMP-BS} to update channel coefficients $\bm\Omega^{(u)}[k]$ and $ \bm b^{(u)}_{l}$, as well as the path parameters $\hat{\bm\varphi}_{\hat{l}}^{(u)}$. 
		\item[] \quad \quad 2. \textit{Dictionary update stage}: Fixing coefficients, update dictionaries as follows
		\item[] \quad \quad \quad \textbf{While} \textit{not converge} \textbf{do}
		\item[] \quad \quad \quad \quad Update $\bD_{\R,1}$ using \eqref{equ:DR1_update1} and update $\bD_{\T,1}$ similarly,
		\item[] \quad \quad \quad \quad Update $\bD_{\R,2}$ using \eqref{equ:DR2_update2} and update $\bD_{\T,2}$ similarly.	
		%		\item[] \quad \quad \quad \quad
		%		Normalize each column of $ \bD_{\T,1}\left(\bD_{\T,2} \odot \bA_{\T,k}^{\rm v}  \right)	$ and $ \bD_{\R,1}\left(\bD_{\R,2} \odot \bA_{\R,k}^{\rm v}  \right)	$.
		\item[] \quad \quad \quad \textbf{end while} 
		\item[] \textbf{end while} 
		\item \textbf{Output:} The optimal dictionaries $\bD_{\T,1},\bD_{\T,2},\bD_{\R,1},\bD_{\R,2}$.
	\end{itemize}  
\end{algorithm} 

\begin{figure}[h!] 
	\centering 
	\includegraphics[width=0.8\textwidth]{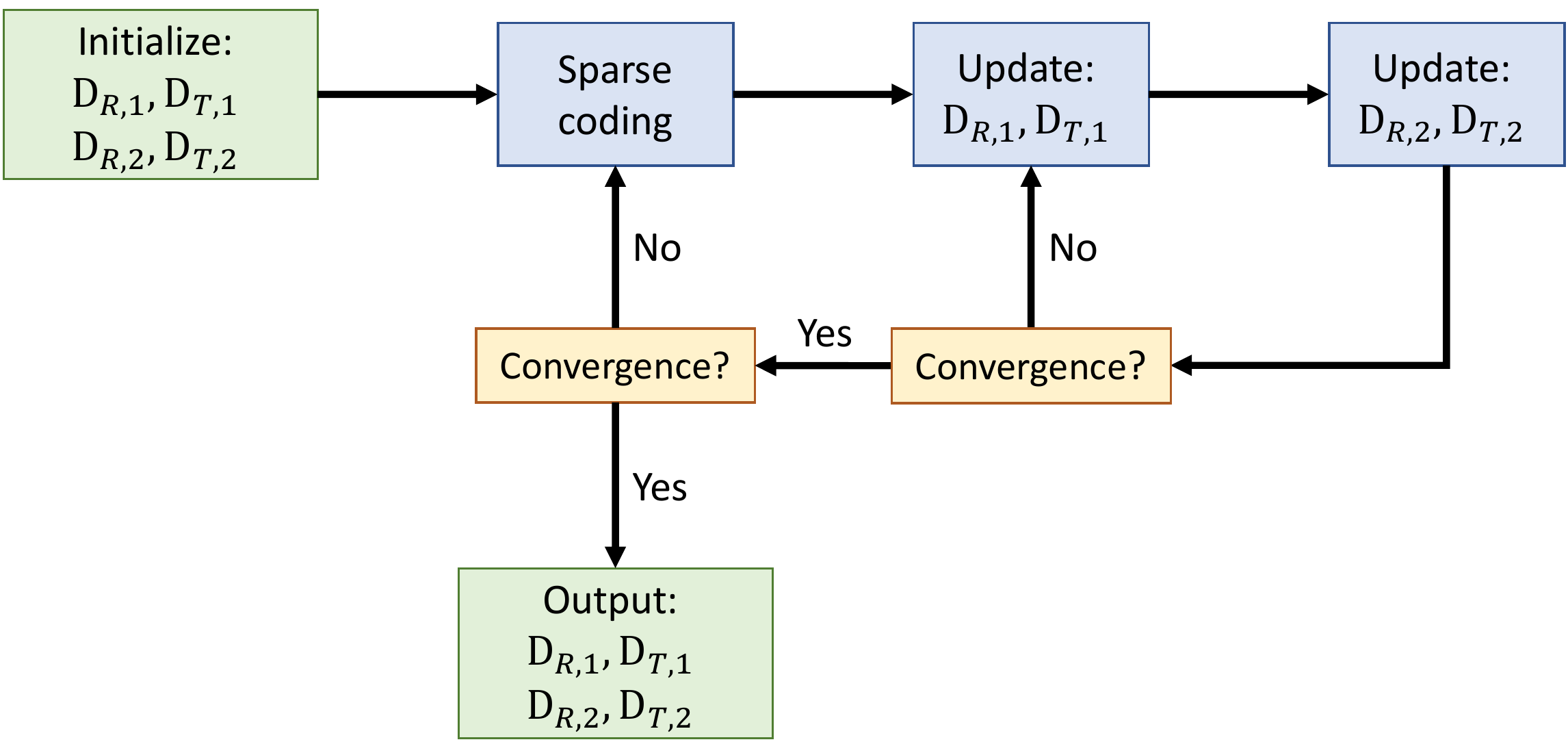}
	\caption{Flow diagram of the dictionary learning stage.}\label{fig:flow}
\end{figure}

The overall procedure of the proposed DL scheme is summarized in Algorithm \ref{alg:DLHW}, and the flow diagram is represented in Fig.~\ref{fig:flow}.

Once the hardware impairments related dictionaries $\bD_{\R,1},\bD_{\R,2},\bD_{\T,1},\bD_{\T,2}$ are obtained, the overall RX and TX sparsifying dictionaries $\breve{\bA}^{\rm v}_{\R,k}$ and $\breve{\bA}^{\rm v}_{\T,k}$ can be constructed at each subcarrier to incorporate the beam squint impact as in \eqref{equ:ARkv}, and then used for subsequent online compressive channel estimation, which is expected to help reduce the training overhead significantly.

\subsection{Convergence and complexity analysis}
% The alternating optimization involves two stages. For sparse coding stage, as explained in \cite{KSVD}, when the sparsity level of signals is small related to its dimension, OMP based pursuit algorithms are known to perform well, i.e., obtaining the best approximation of signals and satisfying the sparsity constraint. Therefore, the proposed OMP based Algorithm \ref{alg:DA-OMP-BS} is able to decrease the value of the objective function during each sparse coding stage.
%While for the dictionary update, an additional reduction or no change in the mismatch error is guaranteed at each sub-stage by using gradient descent with backtracking line search. Therefore, the alternating steps for optimizing the DL problem can guarantee a monotonic decrease in the objective function and then ensure the convergence to a local minimum. 
%

The iterative refinement between the sparse coding and the dictionary update stages is a decreasing process of the objective function in \eqref{equ:DL2}, which takes positive values. In other words, we are minimizing a function bounded by zero and its domain is closed. Therefore, the convergence is guaranteed, and the system will be able to learn the hardware impairments.

As for the complexity, since the DL phase can be implemented offline, the complexity involved in the dictionary update stage does not increase the overall complexity of the online sparse coding stage. Therefore, we compare the computational overhead of the proposed DA-OMP-BS with those of TD-OMP \cite{VenAlkPre:Channel-Estimation-for-Hybrid:17},
in terms of complex multiplication operations for each iteration. 
For ease of comparison, we set the numbers of AoA/AoD/delay grids to be the same $\Kr=\Kt=L_{\rm v} = S$.
For sparse coding with DA-OMP-BS, the parameters can be estimated by iterative one-dimensional search over each parameter space, and thus the complexity order per iteration is $\cO((\Nr+\Nt +\Nc)S\Nsa)$. For TD-OMP, the joint search over the 3D parameter space induces a much higher complexity order of $\cO\big(\Nr\Nt +S^3\Nc\Nsa\big)$.

\section{Numerical Results}\label{sec:simulations} 
In this section, we present numerical results  to demonstrate the effectiveness of the proposed dictionary learning and channel estimation algorithms for a realistic hybrid wideband mmWave MIMO system under both hardware impairments and beam squint.  

%Paramaters of the TX/RX
The TX and RX are equipped with ULAs with half-wavelength spacing, i.e., $\dr=\dt=\lambdac/2$, and $\Nt=32$, $\Nr=8$. Regarding the RF chains, $\Lt=\Lr=\Ns=2$.  The number of OFDM subcarriers is $\Nc=64$ and $\Ts=1~ ns$. 

%Paramaters of the channel
The channels are generated based on \eqref{equ:Hk_two_cases} with $L=6$ multipath components, as in typical indoor scenarios. The pulse shaping function $p(t)$ is assumed to be a raised-cosine filter with roll-off factor of $\beta=0.25$. The  angle of arrival $\phi_l$ and angle of departure $\theta_l$ are uniformly distributed in $[-\pi, \pi]$, while the delays
 are uniformly distributed in $[0, 16T_{\rm s}]$. The gains are generated following a complex Gaussian distribution with the variance adjusted to achieve a specific SNR for the evaluation of the channel estimation strategies. 
%The SNR is set to 0 dB for DL and channel estimation. 
For the parameters of hardware impairments, as in  \cite{EbeEscBie:Investigations-on-antenna-array:16,xie2020dictionary,ding2018dictionary}, the maximum gain and phase error variances for each antenna element are set as $5\%$ and $20^\circ\pi/180^\circ$ respectively. Finally, the mutual coupling coefficients among antennas are within $[0.01,0.4]$ and the antenna location errors are assumed to be uniformly distributed between $[-0.1\lambdac,0.1\lambdac]$. 

%Parameters and algs for channel estimation and algs
We evaluate the performance of our proposed wideband channel estimation algorithm, DA-OMP-BS, in addition to two baseline algorithms described in prior work, denoted as TD-OMP \cite{VenAlkPre:Channel-Estimation-for-Hybrid:17} and  WB-ADMM \cite{Vlachos2019}. The sizes for the discrete angle and delay grids are set as $\Kt=2\Nt$, $\Kr=2\Nr$ and $L_{\rm v}=2\Nc$ for all the different algorithms evaluated in the simulations. Note, however, that the proposed Algorithm \ref{alg:DA-OMP-BS} is of much lower complexity, so that finer angle and delay grids could have been used to improve parameter estimation accuracy.
The performance of the different channel estimation strategies is evaluated in conjunction with different strategies for building the sparsifying dictionaries. In particular, we will consider the dictionaries obtained with our proposed approach DLHWBS, a dictionary constructed from overcomplete ideal array response matrices (IARM) $\bA_{\R,k}^{\rm v}$ and $\bA_{\T,k}^{\rm v}$ assuming no any hardware impairments, the general SeDL algorithm proposed in \cite{xie2020dictionary} which learns a frequency-flat combined dictionary for hardware impairments, and the ideal dictionary built from the known impairments and the corresponding array response vectors.

%Paramaters for DL
During the DL phase, the number of OFDM symbols for training is set as $500$, and a spreading factor $10$ is used to increase the effective SNR by 10 dBs.
For the optimization of DLHWBS,  a revised version of the dictionary initialization algorithm (DIA) in \cite{RusDum:An-initialization-strategy-for-the-dictionary:14} is used for  initialization of the hardware impairments. Specifically, the original DIA algorithm \cite{RusDum:An-initialization-strategy-for-the-dictionary:14} is first applied to initialize the combined dictionary $\bD_{\R}^{\text{init}}\teq \bD_{\R,1}(\bD_{\R,2}\odot\bA_{\R,k}^{\text{v}})$. Then we set the initial values of the antenna spacing errors as zeros, i.e., $\bm\epsilon_{\R}=\b0$ and $\bD_{\R,2}=\bm 1_{\Nr\times \Kr}$, and thus the initialization of gain, phase and coupling matrix is obtained by $\bD_{\R,1} = \bD_{\R}^{\text{init}} ( \bA_{\R,k}^{\text{v}})^{\dag}$. 

\begin{figure}[h!] 
	\centering 
	\includegraphics[width=0.35\textwidth, trim=115px 225px 125px 225px, clip]{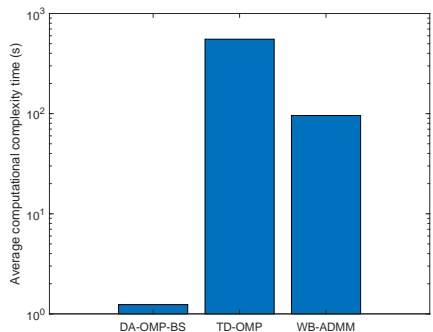}
	\caption{Average computation complexity time of each evaluated channel estimation method.}\label{fig:Time}
\end{figure}

First we evaluate the average computational complexity time for each channel estimation algorithm considering 100 channel realizations.  The results are shown in 
Fig. \ref{fig:Time}, where it can be seen that with the selected grid sizes, fixed for all the algorithms,  our method is around 500 times faster then TD-OMP, and around 100 times faster than WD-ADMM. In the next simulations we will show that despite this significant reduction in complexity, our method always outperforms WD-ADMM for any selection of the dictionary, and, depending on the system parameters, it slightly outperforms or performs similarly to TD-OMP. 

\begin{figure}[h] 
	\centering 
	\subfigure[]{
		\includegraphics[width=0.35\textwidth, trim=115px 225px 125px 225px, clip]{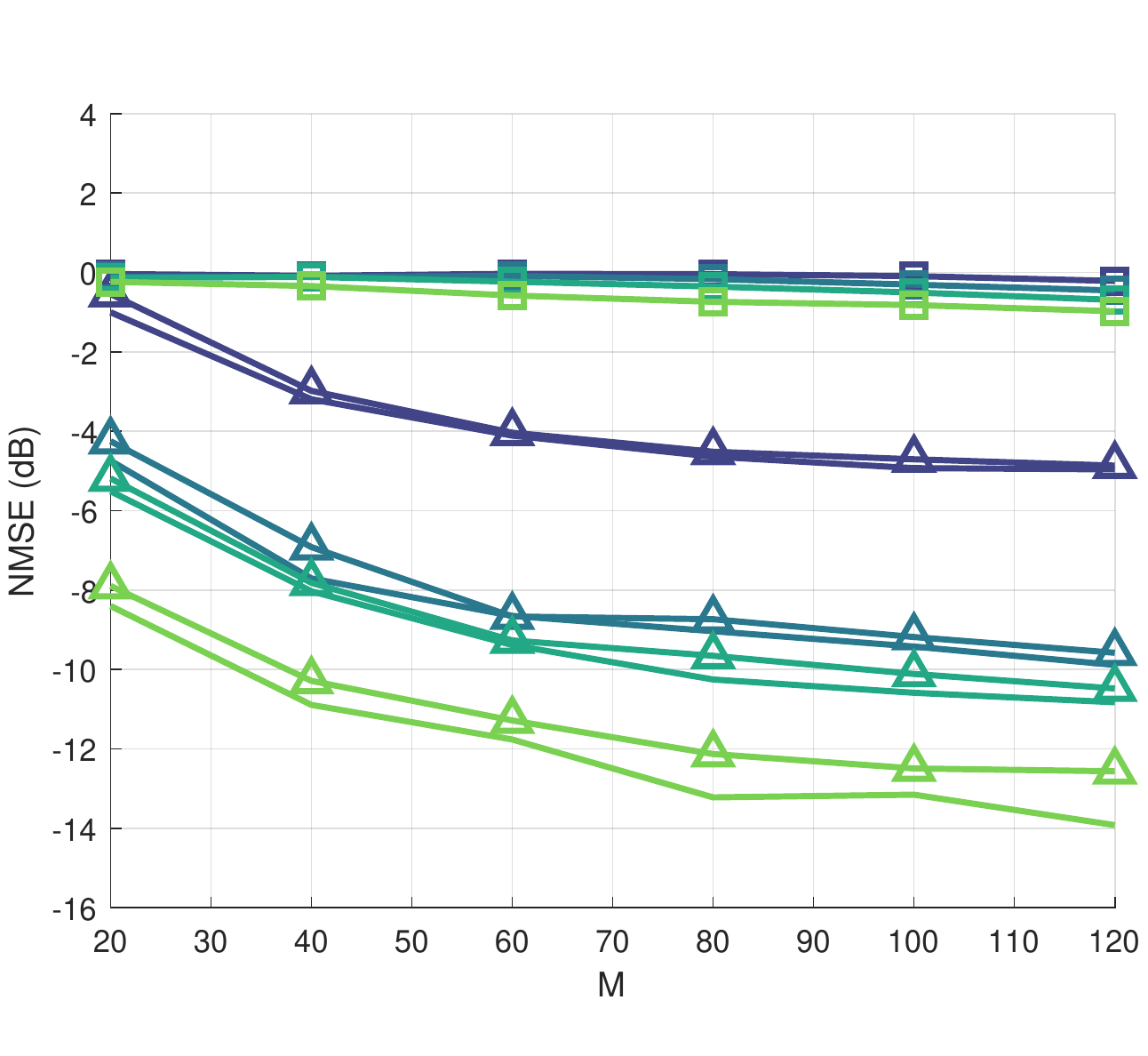}\label{fig:MSE_M}}
	\includegraphics[width=0.25\textwidth, trim=200px 250px 200px 300px, clip]{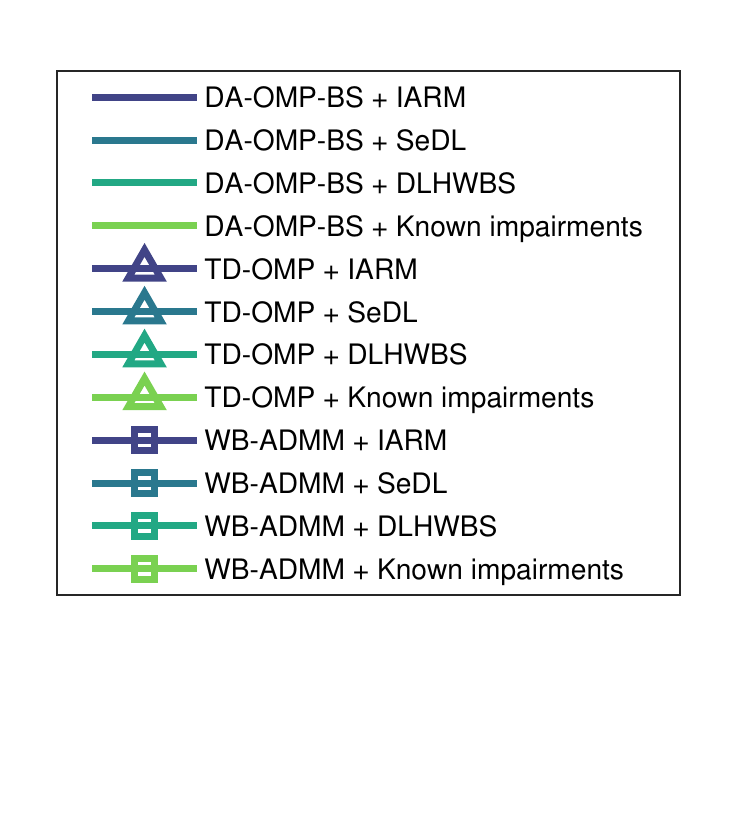}
	\subfigure[]{
		\includegraphics[width=0.35\textwidth, trim=115px 225px 125px 225px, clip]{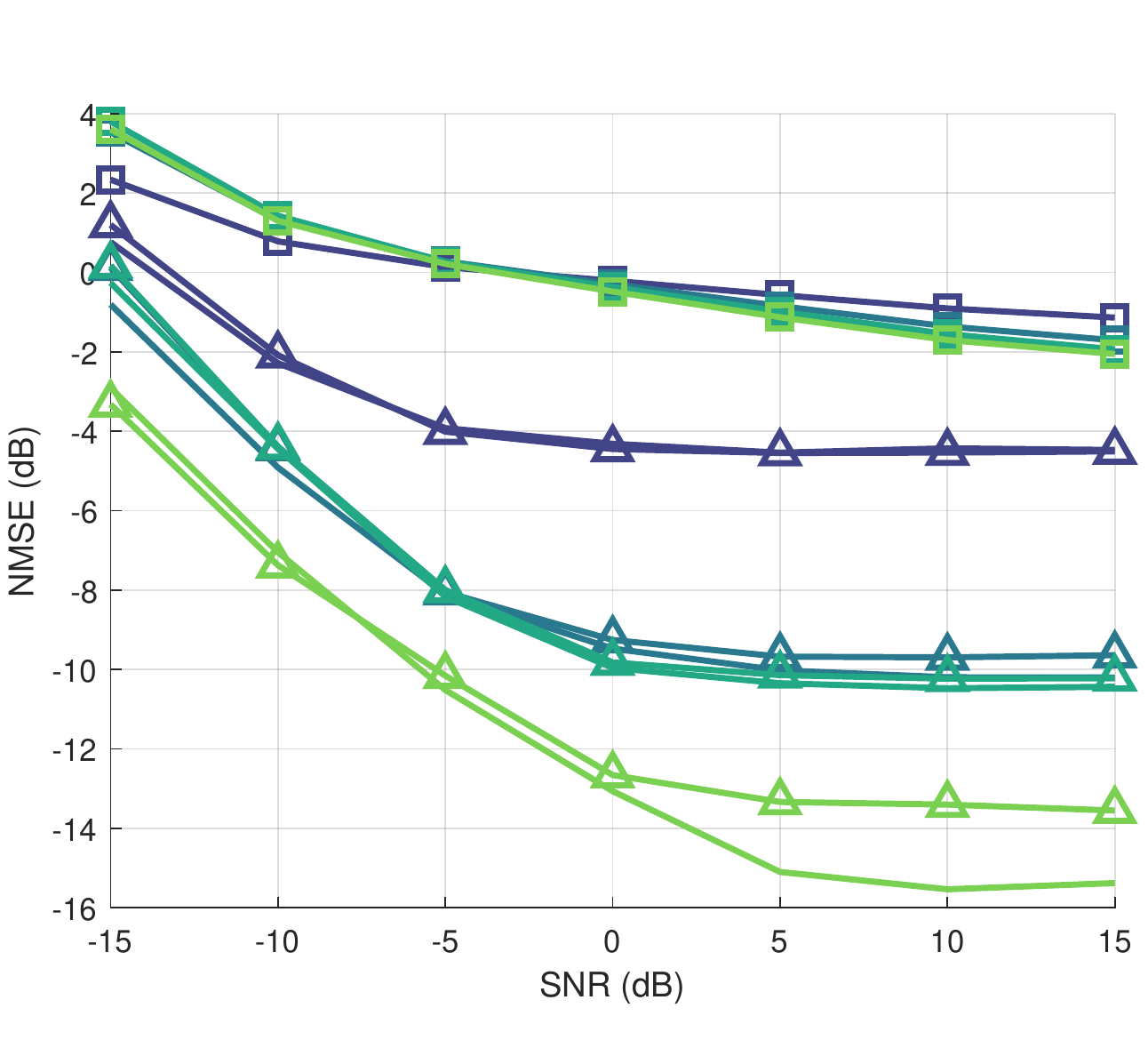}\label{fig:MSE_SNR}}
	\caption{Comparisons of NMSE performance: (a) as a function of the number of training OFDM symbols and SNR set to $0{\rm dB}$; (b)  as a function of the SNR for  $60$ training OFDM symbols.}\label{fig:MSE} 
\end{figure}

Next we evaluate the normalized mean squared error (NMSE) for the different combinations of channel estimation algorithms and sparsifying dictionaries, as a function of the number of training symbols $M$ and the SNR considering 100 channel realizations. Fig.~\ref{fig:MSE}(a)  shows the NMSE when the SNR=0dB and the number of training symbols varies from 20 to 120.
It can be observed that the NMSE reduction provided by DA-OMP-BS when exploiting the ideal dictionary based on known impairments varies from 1 to 1.5 dB. DA-OMP-BS outperforms any other channel estimation algorithm independently of the considered dictionary. Fig.~\ref{fig:MSE}(b) shows the NMSE results as function of the SNR when the number of training symbols $M$ is set to 60. For SNR=0dB, both DA-OMP-BS leads to an NMSE value of -10 dB as TD-OMP but with a cost in complexity 500 times lower. 
Leaving aside the ideal dictionary built from known impairments, the best performing dictionary is  DLHWBS, as it can be seen in both Fig.~\ref{fig:MSE}(a) and Fig.~\ref{fig:MSE}(b). DLHWBS outperforms SeDL because the latter attempts to learn a frequency-flat dictionary for all subcarriers under both hardware impairments and beam squint effect, and assumes all the channels follow the second case of \eqref{equ:Hk_two_cases}. In other words, the distortions on the channel models at side subcarriers are ignored, and thus there already exist some modeling errors when SeDL is applied. These results confirm the effectivenes of our proposed approaches both for channel estimation and dictionary learning. 

\begin{figure}[t] 
	\centering 
	\subfigure[]{
		\includegraphics[width=0.35\textwidth, trim=115px 225px 125px 225px, clip]{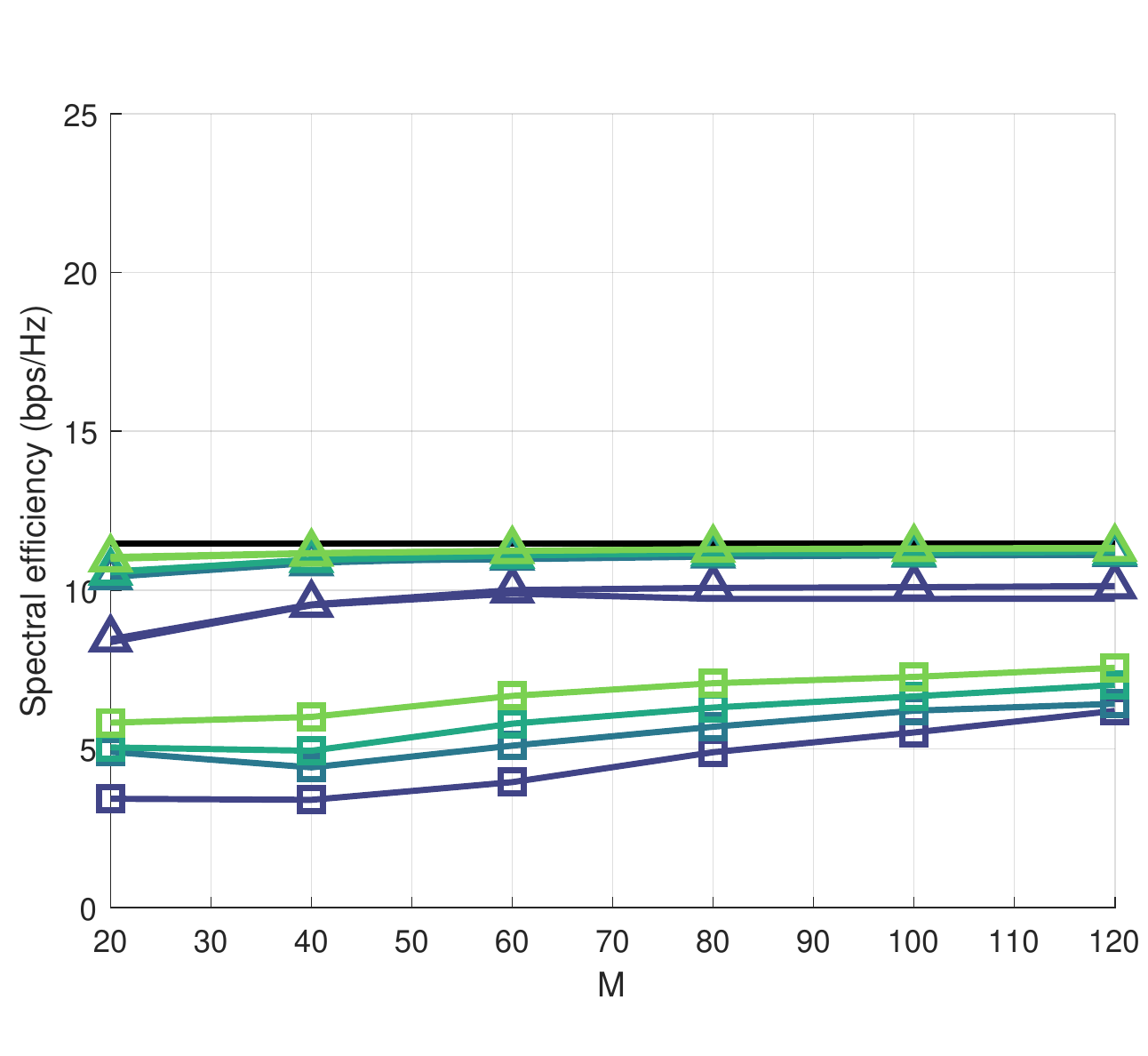}\label{fig:SE_M}}
	\includegraphics[width=0.25\textwidth, trim=200px 250px 200px 300px, clip]{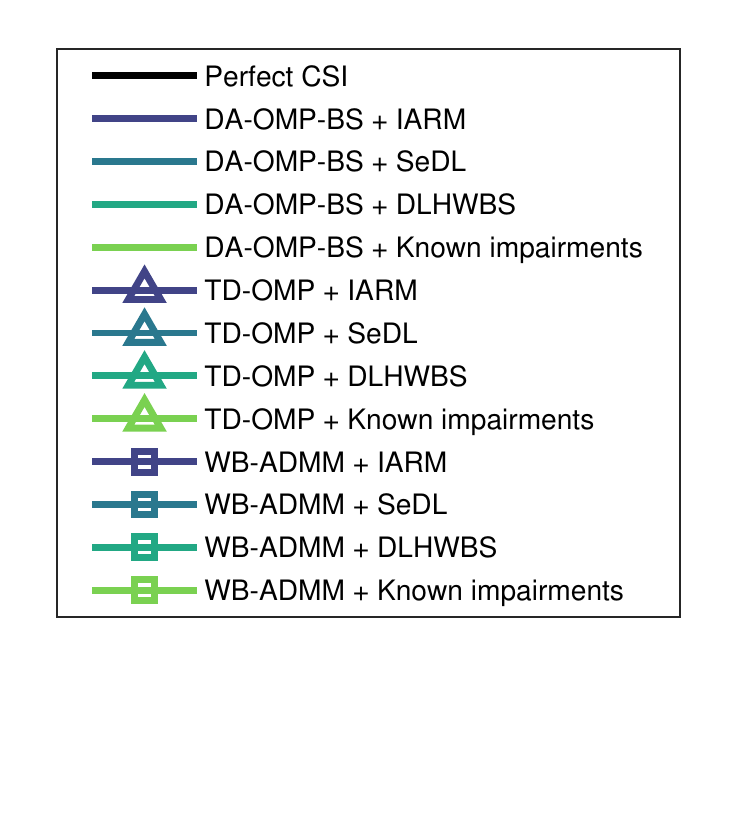}
	\subfigure[]{
		\includegraphics[width=0.35\textwidth, trim=115px 225px 125px 225px, clip]{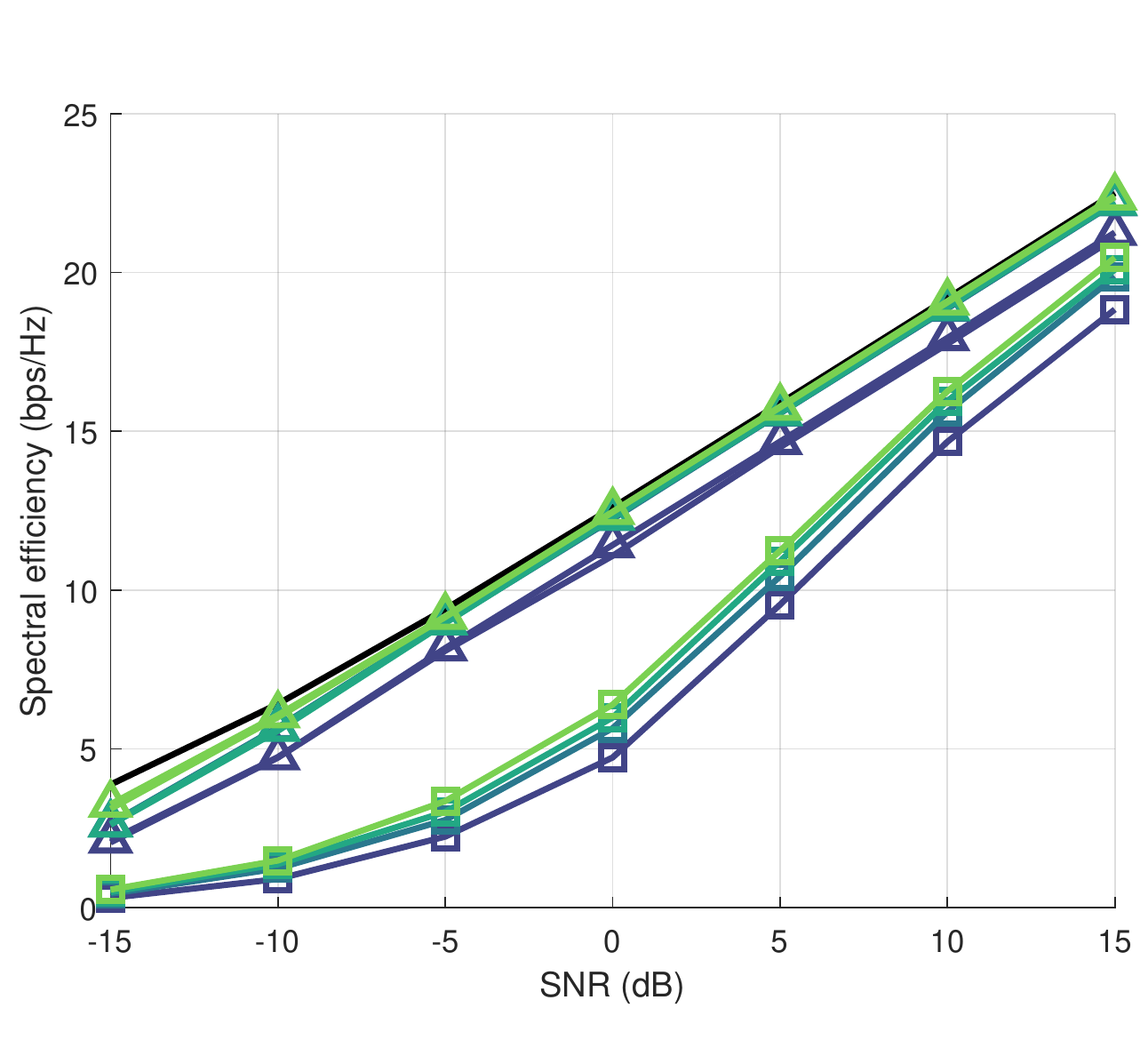}\label{fig:SE_SNR}}
	\caption{Comparison of SE performance: (a) as a function of the number of training OFDM symbols and SNR set to $0{\rm dB}$; (b)  as a function of the SNR for  $60$ training OFDM symbols.}\label{fig:SE} 
\end{figure}

In Fig. \ref{fig:SE}, we also compare the spectral efficiency (SE) performances corresponding to the various sparsifying dictionaries and sparse coding algorithms. As in previous work \cite{xie2020dictionary,rodriguez2018frequency}, the SE is computed by assuming fully-digital precoding and combining using estimates for the $\Ns$ dominant left and right singular vectors of the channel estimates. To be clear, SE is defined as 
\begin{align*} 
\text{SE}=\frac{1}{K}\sum_{k=0}^{K-1}\sum_{n=1}^{\Ns}\log\left(1+\frac{\text{SNR}}{\Ns}\lambda_n( \bH_{\text{eff}}[k])^2\right),
\end{align*} 
where $\bH_{\text{eff}}[k]$ is the effective channel after precoding/combining and $\lambda_n( \bH_{\text{eff}}[k])$ takes the singular values of $ \bH_{\text{eff}}[k]$. For both DA-OMP-BS and conventional TD-OMP, the SE can be significantly increased when exploiting the dictionaries learned with DLHWBS instead of IARM dictionaries. 
Moreover, the performance gap between the proposed DLHWBS algorithm and the case of ideal hardware impairment knowledge is small.

\begin{figure}[h] 
	\centering 
	\subfigure[]{
		\includegraphics[width=0.35\textwidth, trim=105px 225px 125px 225px, clip]{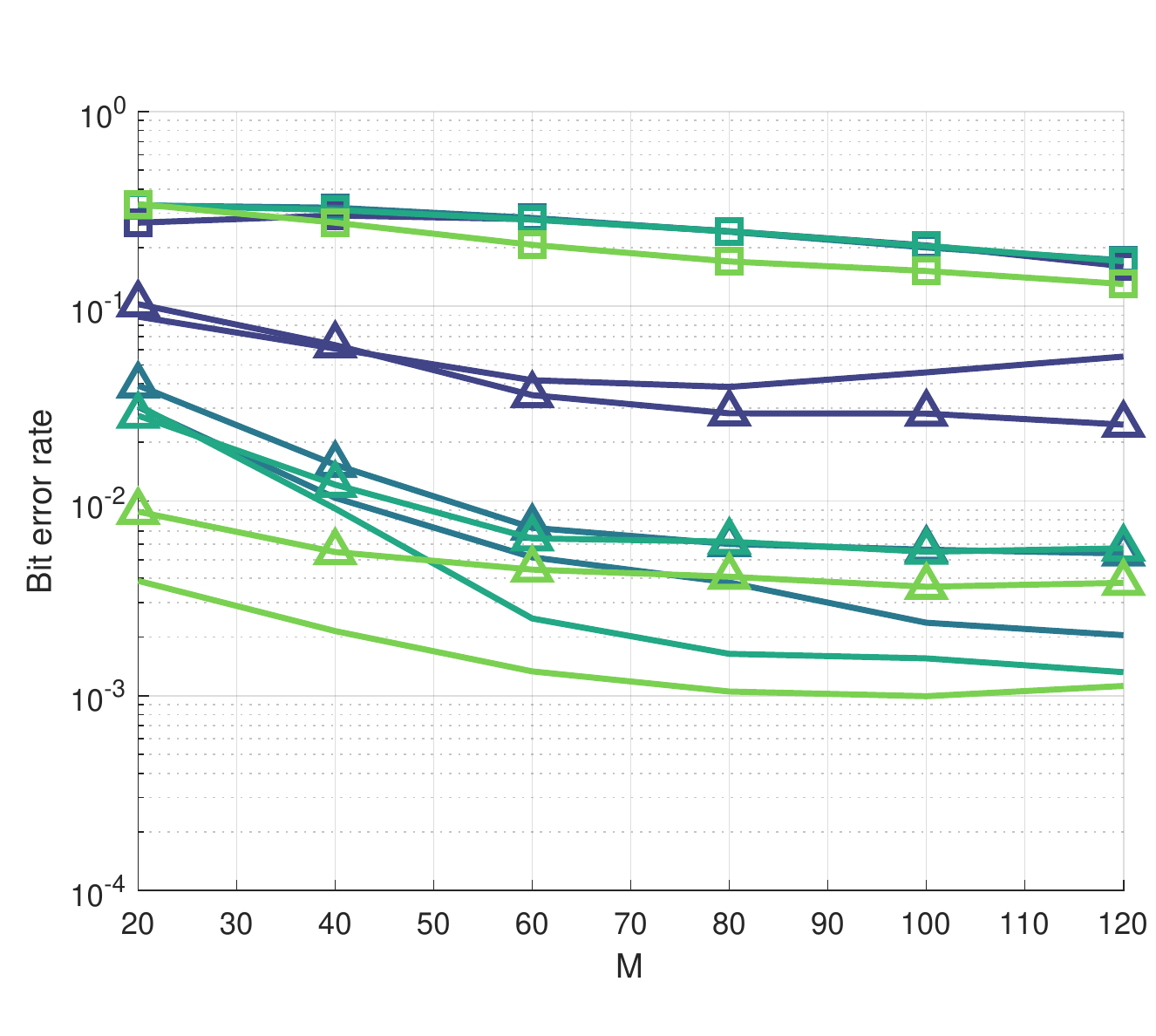}\label{fig:BER_M}}
	\includegraphics[width=0.25\textwidth, trim=200px 250px 200px 300px, clip]{./figures/legend_red.pdf}
	\subfigure[]{
		\includegraphics[width=0.35\textwidth, trim=105px 225px 125px 225px, clip]{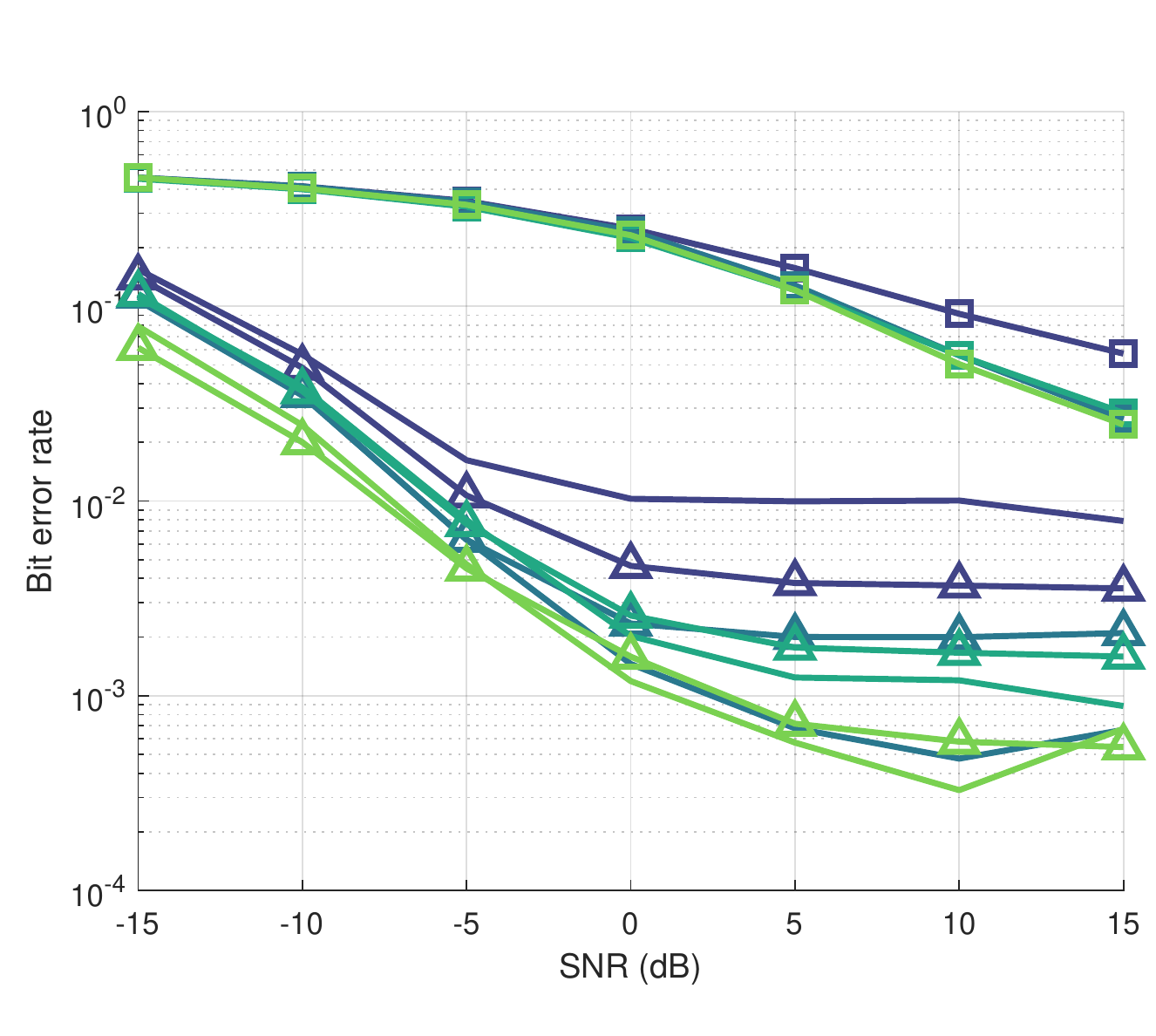}\label{fig:BER_SNR}}
	\caption{Comparisons of BER performance: (a) as a function of the number of training OFDM symbols and SNR set to $0{\rm dB}$; (b)  as a function of the SNR for  $60$ training OFDM symbols.}\label{fig:BER} 
\end{figure}

Finally, we compute the BER when considering the quadrature phase shift keying (QPSK) modulation and minimum mean squared error (MMSE) detection. Fig. \ref{fig:BER}.
shows  how better channel estimates  translate into lower BER for DA-OMP-BS combined with DLHWBS.
The average BER performance gap between the proposed DLHWBS and the ideal case of known impairments is less than 0.5 dB. This reaffirms the effectiveness of our proposed algorithm for hybrid wideband channel under hardware impairments and beam squint.

\section{Conclusions}
In this manuscript, we  derived a general channel model for MIMO systems  by explicitly considering the combined impact of hardware impairments, pulse shaping/filtering effects and beam squint. The resulting model   is an extension of existing MIMO channel models with beam squint. Based on this general channel model, we formulated a DL problem to obtain the sparsifying dictionaries for channel representation that account for hardware impairments. The effect of beam squint is considered, but it is not part of the learning process, since it can be mathematically modeled. We also proposed a novel compressive channel estimation algorithm under beam squint, which exploits the structure of the beam squint at different subcarriers to  facilitate the estimation of the channel parameters at  a much lower complexity. Numerical results have demonstrated the effectiveness of the proposed DL and channel estimation strategies and the significant gains with respect to the strategies proposed in prior work. 

%\newpage
 
\section*{Appendix I} \label{sec:appendix}
\setcounter{subsection}{0}
\section*{Proof of equation \eqref{equ:yk_predef}}

Let the delayed filter be
\begin{align} 
r(t) = p(t-\tau). 
\end{align}
If we have a set of measurements ${\bf r} = [r(0),r(\Ts), \ldots, r( (K -1)\Ts)]^T$, the $k$-th entry after DFT can be computed as
\begin{align}
\hat{r}[k] = \sum_{d=0}^{K-1} [\br]_d\cdot e^{-\j\frac{2\pi kd}{K}}.
\end{align}
Equivalently, we can use the continuous interpretation of the DFT to represent this as
\begin{align}
\hat{r}[k] =
\mathcal{F}\Big(\sum_d \delta_{dT_{\rm s}}(p*\delta_{\tau}) \Big)[\Delta f_k],
\end{align} 
where $\sum_d \delta_{dT_{\rm s}} = \sum_{d=-\infty}^{\infty} \delta(t-d\Ts)$ is the sampling function with period $\Ts$ and $\delta_{\tau} \teq \delta(t-\tau)$. Moreover, $*$ denotes convolution operation and $\cF(\cdot)[\Delta f_k]$ is the Fourier transform evaluated for the frequency difference $\Delta f_k$.
Next we can make use of Fourier product and convolution identities to get to
\begin{align}
\hat{r}[k] =
\Big(\mathcal{F}\big(\sum_d\delta_{dT_{\rm s}}\big)* \big(\mathcal{F}(p)\mathcal{F}(\delta_{\tau}) \big) \Big)[\Delta f_k].
\end{align}
Using the Dirac comb Fourier identity formula we reach
\begin{align}
\hat{r}[k] =
\Big(\big(\frac{1}{T_{\rm s}}\sum_i\delta_{i/T_{\rm s}}\big)*\big(\mathcal{F}(p)\mathcal{F}(\delta_{\tau})\big) \Big)[\Delta f_k].
\end{align}
Now we can express the convolution in terms of an integral
\begin{align}
\hat{r}[k] = \frac{1}{T_{\rm s}}\int \mathcal{F}(p)[f]  {e^{-\j2\pi f\tau } }\sum_i\delta_{i/T_{\rm s}}(\Delta f_k-f) df,
\end{align}
and this can be solved by evaluating the Dirac comb as follows
\begin{align}\label{equ:rk}
\hat{r}[k] = \sum _i \frac{\mathcal{F}(p)[\Delta f_k+i/T_{\rm s}]}{T_{\rm s}}e^{-\j 2\pi (\Delta f_k+i/T_{\rm s})\tau } = \Big(\sum _i \frac{\mathcal{F}(p)[\Delta f_k+i/T_{\rm s}]}{T_{\rm s}}e^{-\j2\pi i\tau/T_{\rm s} }\Big)e^{-\j2\pi\Delta f_k\tau }.
\end{align}
Note that since $\mathcal{F}(p)$ is bounded, this sum only includes a few terms.
Let us define the delay-frequency distortion as
\begin{align} \label{equ:g_k_tau}
g(k, \tau) = \sum_{k'} \frac{\mathcal{F}(p)[\Delta f_k+k'/T_{\rm s}]}{T_{\rm s}}e^{-\j2\pi i\tau/T_{\rm s} },
\end{align} 
and then \eqref{equ:rk} can be simplified to
\begin{align}
\hat{r}[k] = g(k, \tau)e^{-\j2\pi\Delta f_k\tau}.
\end{align}
If we assume the domain of $\mathcal{F}(p)$ to be in $[-\frac{1+\beta}{2T_s}, +\frac{1+\beta}{2T_s}]$ with $\beta\in[0, 1]$, then we have the expression for $g(k, \tau) $ as
\begin{align}
g(k, \tau) = \left\lbrace
\begin{array}{cl}
\frac{\mathcal{F}(p)[\Delta f_k]}{T_{\rm s}}+\frac{\mathcal{F}(p)[\Delta f_k-1/T_{\rm s}]}{T_{\rm s}}e^{\j2\pi\tau/T_{\rm s}} & \text{if} \ \ 2\Delta f_k > (1-\beta)/T_{\rm s} \\
\frac{\mathcal{F}(p)[\Delta f_k]}{T_{\rm s}} & \text{if}\ \ 2|\Delta f_k| \leq (1-\beta)/T_{\rm s} \\
\frac{\mathcal{F}(p)[\Delta f_k]}{T_{\rm s}}+\frac{\mathcal{F}(p)[\Delta f_k+1/T_{\rm s}]}{T_{\rm s}}e^{-\j2\pi\tau/T_{\rm s}} & \text{if}\ \ 2\Delta f_k < -(1-\beta)/T_{\rm s}
\end{array}
\right.
\end{align} 
In the case of a raised cosine filter with parameters $T=T_{\rm s}$ and $\beta\in[0, 1]$, it is straightforward to prove that
\begin{align} 
g(k, \tau) = \left\lbrace
\begin{array}{cl}
\frac{1}{2}(1+e^{\j\frac{2\pi\tau}{T_{\rm s} }}+(1-e^{\j\frac{2\pi\tau}{T_{\rm s}}})\cos(\frac{\pi T_{\rm s}}{\beta}(|\Delta f_k| - \frac{1-\beta}{2T_{\rm s}}))) & \text{if}\ \ 2\Delta f_k > (1-\beta)/T_{\rm s} \\
1 & \text{if}\ \ 2|\Delta f_k| \leq (1-\beta)/T_{\rm s} \\
\frac{1}{2}(1+e^{-\j\frac{2\pi\tau}{T_{\rm s} }}+(1-e^{-\j\frac{2\pi\tau}{T_{\rm s}}})\cos(\frac{\pi T_{\rm s}}{\beta}(|\Delta f_k| - \frac{1-\beta}{2T_{\rm s}}))) & \text{if}\ \ 2\Delta f_k < -(1-\beta)/T_{\rm s}
\end{array}
\right.
\end{align}
 
\section*{Appendix II}
\section*{Derivative of the objective function in \eqref{equ:DR1_obj2} with respect to $\bD_{\rm R,1}$}
 
First, we calculate the derivative of the first sum term in \eqref{equ:DR1_obj2} with respect to $\bD_{\rm R,1}$. For $u\in\cI(\Nsa), k\in\cKcen$, we have
\begin{align} \label{equ:derivative_DR1_1}
 \frac{\partial J_k^{(u)}}{\partial \bD_{\R,1}}  =   - \overline{( \hat{\bH}^{(u)}[k]- \bD_{\R,1}\bX_{\R,1}^{(u)}[k]  )}(\bX_{\R,1}^{(u)}[k])^T.
\end{align}

Next, we calculate the derivative of the second sum term in \eqref{equ:DR1_obj2} with respect to $\bD_{\rm R,1}$. Recalling the chain rule, we can express the Jacobian matrix of the second sum term with respect to $\bD_{\rm R,1}$ as, $\forall u\in\cI(\Nsa), k\in\cKside$,
\begingroup
\allowdisplaybreaks
\begin{align}
	\frac{\partial J_k^{(u)}}{\partial \vec( \bD_{\R,1})^T}  & = 
	\frac{\partial J_k^{(u)}}{\partial \vec(\sum_{\hat{l}=1}^{\hat{L}} \hat{\alpha}_{\hat{l}} e^{-\j 2\pi\Delta f_k\hat{\tau}_{\hat{l}}}  {\bf G}_k(\hat{\tau}_{\hat{l}}, \hat{\phi}_{\hat{l}}, \hat{\theta}_{\hat{l}})\odot  \bD_{\R,1}\bX_{\R,1,\hat{l}}^{(u)}[k] )^T} \notag\\
	&  \cdot \frac{\partial \vec(\sum_{\hat{l}=1}^{\hat{L}} \hat{\alpha}_{\hat{l}} e^{-\j 2\pi\Delta f_k\hat{\tau}_{\hat{l}}}  {\bf G}_k(\hat{\tau}_{\hat{l}}, \hat{\phi}_{\hat{l}}, \hat{\theta}_{\hat{l}})\odot  \bD_{\R,1}\bX_{\R,1,\hat{l}}^{(u)}[k] )}{\partial \vec( \bD_{\R,1}\bX_{\R,1,\hat{l}}^{(u)}[k])^T} 
	 \cdot \frac{\partial \vec\big(\bD_{\R,1}\bX_{\R,1,\hat{l}}^{(u)}[k]\big)}{\partial \vec(\bD_{\R,1})^T} \notag\\
	& = -\vec\Big(\overline{\hat{\bH}^{(u)}[k]} - \sum_{\hat{l}=1}^{\hat{L}} \overline{\hat{\alpha}_{\hat{l}} e^{-\j 2\pi\Delta f_k\hat{\tau}_{\hat{l}}}  {\bf G}_k(\hat{\tau}_{\hat{l}}, \hat{\phi}_{\hat{l}}, \hat{\theta}_{\hat{l}})\odot \bD_{\R,1}\bX_{\R,1,\hat{l}}^{(u)}[k]} \Big)^T \notag\\
	& \cdot \sum_{\hat{l}=1}^{\hat{L}} \hat{\alpha}_{\hat{l}} e^{-\j 2\pi\Delta f_k\hat{\tau}_{\hat{l}}} \diag\big\{\vec({\bf G}_k(\hat{\tau}_{\hat{l}}, \hat{\phi}_{\hat{l}}, \hat{\theta}_{\hat{l}})) \big\}   \big( (\bX_{\R,1,\hat{l}}^{(u)}[k])^T\kron \bI_{\Nr}\big).
\end{align}
\endgroup
With the Jocobian matrix, we can expressive the derivative of the second sum term in \eqref{equ:DR1_obj2} with respect to $\bD_{\rm R,1}$ as follows
\begingroup
\allowdisplaybreaks
\begin{align}\label{equ:derivative_DR1_2}
	& \frac{\partial J_k^{(u)}}{\partial \bD_{\R,1}}  = \text{unvec}\Big\{\frac{\partial J_k^{(u)}}{\partial \vec( \bD_{\R,1})} \Big\} = \text{unvec}\Big\{ -\sum_{\hat{l}=1}^{\hat{L}} \hat{\alpha}_{\hat{l}} e^{-\j 2\pi\Delta f_k\hat{\tau}_{\hat{l}}}  \big(  \bX_{\R,1,\hat{l}}^{(u)}[k] \kron \bI_{\Nr}\big) \notag\\
	& \kern 10pt \cdot \diag\big\{\vec({\bf G}_k(\hat{\tau}_{\hat{l}}, \hat{\phi}_{\hat{l}}, \hat{\theta}_{\hat{l}})) \big\}  \cdot \vec\Big(\overline{\hat{\bH}^{(u)}[k]} - \sum_{\hat{l}=1}^{\hat{L}} \overline{\hat{\alpha}_{\hat{l}} e^{-\j 2\pi\Delta f_k\hat{\tau}_{\hat{l}}}  {\bf G}_k(\hat{\tau}_{\hat{l}}, \hat{\phi}_{\hat{l}}, \hat{\theta}_{\hat{l}})\odot \bD_{\R,1}\bX_{\R,1,\hat{l}}^{(u)}[k]} \Big) \Big\}
	\notag \\
	& = -\sum_{\hat{l}=1}^{\hat{L}} \hat{\alpha}_{\hat{l}} e^{-\j 2\pi\Delta f_k\hat{\tau}_{\hat{l}}} \Big[\Big(\overline{\hat{\bH}^{(u)}[k]} - \sum_{\hat{l}=1}^{\hat{L}} \overline{\hat{\alpha}_{\hat{l}} e^{-\j 2\pi f_k\hat{\tau}_{\hat{l}}}  {\bf G}_k(\hat{\tau}_{\hat{l}}, \hat{\phi}_{\hat{l}}, \hat{\theta}_{\hat{l}})\odot \bD_{\R,1}\bX_{\R,1,\hat{l}}^{(u)}[k]} \Big) \odot  {\bf G}_k(\hat{\tau}_{\hat{l}}, \hat{\phi}_{\hat{l}}, \hat{\theta}_{\hat{l}})\Big]\notag\\
	&\kern 40pt \cdot \big(  \bX_{\R,1,\hat{l}}^{(u)}[k] \big)^T.
\end{align} 
\endgroup 
Then combining the derivatives in $\eqref{equ:derivative_DR1_1}$ for $k\in\cKcen$ and the derivatives in $\eqref{equ:derivative_DR1_2}$ for $k\in\cKside$, we can obtain the final derivative of the objective function in \eqref{equ:DR1_obj2} with respect to $\bD_{\rm R,1}$, i.e.,
\begin{align} 
& \frac{\partial J }{\partial \bD_{\R,1}}  =   - \sum_{u\in\cI(\Nsa)}\sum_{k\in\cKcen} \overline{( \hat{\bH}^{(u)}[k]- \bD_{\R,1}\bX_{\R,1}^{(u)}[k]  )}(\bX_{\R,1}^{(u)}[k])^T  
 - \sum_{u\in\cI(\Nsa)}\sum_{k\in\cKside} \sum_{\hat{l}=1}^{\hat{L}} \hat{\alpha}_{\hat{l}} e^{-\j 2\pi\Delta f_k\hat{\tau}_{\hat{l}}} \notag \\ 
 & \cdot \Big[\Big(\overline{\hat{\bH}^{(u)}[k]} - \sum_{\hat{l}=1}^{\hat{L}} \overline{\hat{\alpha}_{\hat{l}} e^{-\j 2\pi\Delta f_k\hat{\tau}_{\hat{l}}}  {\bf G}_k(\hat{\tau}_{\hat{l}}, \hat{\phi}_{\hat{l}}, \hat{\theta}_{\hat{l}})\odot \bD_{\R,1}\bX_{\R,1,\hat{l}}^{(u)}[k]} \Big) \odot  {\bf G}_k(\hat{\tau}_{\hat{l}}, \hat{\phi}_{\hat{l}}, \hat{\theta}_{\hat{l}})\Big]  \big(  \bX_{\R,1,\hat{l}}^{(u)}[k] \big)^T.
\end{align}

\vspace{-10mm}
\section*{Appendix III}
\section*{Derivative of the objective function in \eqref{equ:DR2_obj2} with respect to $\bm\epsilon_{\rm R}$}
 First, we calculate the derivative of the first sum term in \eqref{equ:DR2_obj2} with respect to $\bm\epsilon_{\R}$. 
Note that the gradient of any element of $\bD_{\R,2}$ with respect to the antenna location error $\epsilon_{\R,m}$ can be expressed as
\begin{align}
\frac{\partial [\bD_{\R,2}]_{m,n}}{\partial \epsilon_{\R,m}} = \frac{\partial e^{-\j2\pi \fc \epsilon_{\R,m} \cdot \sin(\phi_n^{\rm v})/c   } }{\partial \epsilon_{\R,m}}  
=  [\bD_{\R,2}]_{m,n} \cdot \frac{-\j2\pi \fc \sin(\phi_n^{\rm v})}{c}.
\end{align} 
Therefore, we have the gradient of $J_k^{(u)}$ with respect to $\bm\epsilon_{\R}$ as 
\begin{align} \label{equ:epsilon_update1}
\frac{\partial J_k^{(u)}}{\partial \bm\epsilon_\R} = 2\cR\left\{ 
\left\{ \frac{\partial J_k^{(u)}}{\partial  \bD_{\R,2} } \odot \bD_{\R,2} \right\}\cdot \frac{-\j2\pi \fc \sin(\bm\phi^{\rm v})}{c}\right\},
\end{align}

We next calculate the derivative of the second sum term in \eqref{equ:DR2_obj2} with respect to $\bm\epsilon_{\R}$. 
Similar to Appendix II, using the chain rule of Jacobian matrix, we have, $\forall u\in\cI(\Nsa), k\in\cKside$,  
\begingroup
\allowdisplaybreaks
\begin{align}
&\frac{\partial J_k^{(u)}}{\partial   \bm\epsilon_{\R}^T}   = 2\cR\Bigg\{
\frac{\partial J_k^{(u)}}{\partial \vec\Big(\sum_{\hat{l}=1}^{\hat{L}} \hat{\alpha}_{\hat{l}} e^{-\j 2\pi\Delta f_k\hat{\tau}_{\hat{l}}}  {\bf G}_k(\hat{\tau}_{\hat{l}}, \hat{\phi}_{\hat{l}}, \hat{\theta}_{\hat{l}})\odot  \big(\bD_{\R,1} \big(\bee_{\R}(\hat{\phi}_{\hat{l}})\odot   {\ba}_{\R,k}(\hat{\phi}_{\hat{l}}) \big) \bX_{\R,2,\hat{l}}^{(u)}[k] \big) \Big)^T} \notag\\
&  \cdot \frac{\partial \vec\Big(\sum_{\hat{l}=1}^{\hat{L}} \hat{\alpha}_{\hat{l}} e^{-\j 2\pi\Delta f_k\hat{\tau}_{\hat{l}}}  {\bf G}_k(\hat{\tau}_{\hat{l}}, \hat{\phi}_{\hat{l}}, \hat{\theta}_{\hat{l}})\odot  \big(\bD_{\R,1} \big(\bee_{\R}(\hat{\phi}_{\hat{l}})\odot   {\ba}_{\R,k}(\hat{\phi}_{\hat{l}}) \big)\bX_{\R,2,\hat{l}}^{(u)}[k] \big) \Big)}{\partial \vec\big(\bD_{\R,1} \big(\bee_{\R}(\hat{\phi}_{\hat{l}})\odot   {\ba}_{\R,k}(\hat{\phi}_{\hat{l}}) \big)\bX_{\R,2,\hat{l}}^{(u)}[k] \big)^T} \notag\\
& \cdot \frac{\partial \vec\big(\bD_{\R,1} \big(\bee_{\R}(\hat{\phi}_{\hat{l}})\odot   {\ba}_{\R,k}(\hat{\phi}_{\hat{l}}) \big)\bX_{\R,2,\hat{l}}^{(u)}[k] \big)}{\partial \vec\big(\bee_{\R}(\hat{\phi}_{\hat{l}})\odot   {\ba}_{\R,k}(\hat{\phi}_{\hat{l}}) \big)^T}  
 \cdot \frac{\partial \vec\big(\bee_{\R}(\hat{\phi}_{\hat{l}})\odot   {\ba}_{\R,k}(\hat{\phi}_{\hat{l}}) \big)}{\partial \vec\big(\bee_{\R}(\hat{\phi}_{\hat{l}})\big)^T } \cdot 
\frac{\partial \vec\big(\bee_{\R}(\hat{\phi}_{\hat{l}})\big) }{\partial   \bm\epsilon_{\R} ^T } \Bigg\}\notag\\
& = 2\cR\Bigg\{-\vec\Big(\overline{\hat{\bH}^{(u)}[k]} - \sum_{\hat{l}=1}^{\hat{L}} \overline{\hat{\alpha}_{\hat{l}} e^{-\j 2\pi\Delta f_k\hat{\tau}_{\hat{l}}}  {\bf G}_k(\hat{\tau}_{\hat{l}}, \hat{\phi}_{\hat{l}}, \hat{\theta}_{\hat{l}})\odot \big(\bD_{\R,1} \big(\bee_{\R}(\hat{\phi}_{\hat{l}})\odot   {\ba}_{\R,k}(\hat{\phi}_{\hat{l}}) \big) \bX_{\R,2,\hat{l}}^{(u)}[k] \big)} \Big)^T \notag\\
& \cdot \sum_{\hat{l}=1}^{\hat{L}} \alpha_{\hat{l}} e^{-\j 2\pi\Delta f_k\hat{\tau}_{\hat{l}}} \diag\big\{\vec({\bf G}_k(\hat{\tau}_{\hat{l}}, \hat{\phi}_{\hat{l}}, \hat{\theta}_{\hat{l}})) \big\} \notag\\
& \cdot \big( (\bX_{\R,2,\hat{l}}^{(u)}[k])^T\kron \bD_{\R,1}\big)
\cdot \diag\big\{\vec({\ba}_{\R,k}(\hat{\phi}_{\hat{l}}) )\big\} \cdot \diag\big\{   \bee_{\R}(\hat{\phi}_{\hat{l}}) \big\} \cdot \frac{-\j2\pi\fc \sin(\hat{\phi}_{\hat{l}})}{c} 
\Bigg\}.
\end{align}
\endgroup
Therefore, we have 
\begingroup
\allowdisplaybreaks
\begin{align}\label{equ:epsilon_update2}
	\frac{\partial J_k^{(u)}}{\partial   \bm\epsilon_{\R}}  & = 2\cR\Bigg\{-\sum_{\hat{l}=1}^{\hat{L}} \hat{\alpha}_{\hat{l}} e^{-\j 2\pi\Delta f_k\hat{\tau}_{\hat{l}}} \cdot
   \frac{-\j2\pi\fc \sin(\hat{\phi}_{\hat{l}})}{c} \cdot  \diag\big\{   \bee_{\R}(\hat{\phi}_{\hat{l}}) \odot {\ba}_{\R,k}(\hat{\phi}_{\hat{l}})   \big\} \notag\\
   &\cdot \big( (\bX_{\R,2,\hat{l}}^{(u)}[k])\kron \bD_{\R,1}^T\big) \cdot  \diag\big\{\vec({\bf G}_k(\hat{\tau}_{\hat{l}}, \hat{\phi}_{\hat{l}}, \hat{\theta}_{\hat{l}})) \big\} 
   \notag\\
   & \cdot \vec\Big(\overline{\hat{\bH}^{(u)}[k]} - \sum_{\hat{l}=1}^{\hat{L}} \overline{\hat{\alpha}_{\hat{l}} e^{-\j 2\pi\Delta f_k\hat{\tau}_{\hat{l}}}  {\bf G}_k(\hat{\tau}_{\hat{l}}, \hat{\phi}_{\hat{l}}, \hat{\theta}_{\hat{l}})\odot \big(\bD_{\R,1} \big(\bee_{\R}(\hat{\phi}_{\hat{l}})\odot   {\ba}_{\R,k}(\hat{\phi}_{\hat{l}}) \big) \bX_{\R,2,\hat{l}}^{(u)}[k] \big)} \Big) 
	\Bigg\} \notag\\
	& = 2\cR\Bigg\{-\sum_{\hat{l}=1}^{\hat{L}} \hat{\alpha}_{\hat{l}} e^{-\j 2\pi\Delta f_k\hat{\tau}_{\hat{l}}} \cdot \frac{-\j2\pi\fc \sin(\hat{\phi}_{\hat{l}})}{c} \cdot \bigg[ \bD_{\R,1}^T  \notag\\
	& \cdot
	  \bigg(\Big(\overline{\hat{\bH}^{(u)}[k]} - \sum_{\hat{l}=1}^{\hat{L}} \overline{\hat{\alpha}_{\hat{l}} e^{-\j 2\pi\Delta f_k\hat{\tau}_{\hat{l}}}  {\bf G}_k(\hat{\tau}_{\hat{l}}, \hat{\phi}_{\hat{l}}, \hat{\theta}_{\hat{l}})\odot \big(\bD_{\R,1} \big(\bee_{\R}(\hat{\phi}_{\hat{l}})\odot   {\ba}_{\R,k}(\hat{\phi}_{\hat{l}}) \big) \bX_{\R,2,\hat{l}}^{(u)}[k] \big)} \Big) \notag\\
	& \odot {\bf G}_k(\hat{\tau}_{\hat{l}}, \hat{\phi}_{\hat{l}}, \hat{\theta}_{\hat{l}}) \bigg) (\bX_{\R,2,\hat{l}}^{(u)}[k])^T \bigg] \odot \big(\bee_{\R}(\hat{\phi}_{\hat{l}}) \odot {\ba}_{\R,k}(\hat{\phi}_{\hat{l}})\big) \Bigg\}.
\end{align}
\endgroup
Then combining the derivatives in $\eqref{equ:epsilon_update1}$ for $k\in\cKcen$ and the derivatives in $\eqref{equ:epsilon_update2}$ for $k\in\cKside$, we can obtain the final derivative of the objective function in \eqref{equ:DR2_obj2} with respect to $\bm\epsilon_{\R}$, i.e.,
\begingroup
\allowdisplaybreaks 
\begin{align} 
\frac{\partial J}{\partial \bm\epsilon_\R} & = \sum_{u\in\cI(\Nsa)}\sum_{k\in\cKcen} 2\cR\left\{ 
\left\{ \frac{\partial J_k^{(u)}}{\partial  \bD_{\R,2} } \odot \bD_{\R,2} \right\}\cdot \frac{-\j2\pi \fc \sin(\bm\phi^{\rm v})}{c}\right\} \notag\\
& - \sum_{u\in\cI(\Nsa)}\sum_{k\in\cKside}  2\cR\Bigg\{\sum_{\hat{l}=1}^{\hat{L}} \hat{\alpha}_{\hat{l}} e^{-\j 2\pi\Delta f_k\hat{\tau}_{\hat{l}}} \cdot \frac{-\j2\pi\fc \sin(\hat{\phi}_{\hat{l}})}{c} \cdot \bigg[ \bD_{\R,1}^T  \notag\\
& \cdot
\bigg(\Big(\overline{\hat{\bH}^{(u)}[k]} - \sum_{\hat{l}=1}^{\hat{L}} \overline{\hat{\alpha}_{\hat{l}} e^{-\j 2\pi\Delta f_k\hat{\tau}_{\hat{l}}}  {\bf G}_k(\hat{\tau}_{\hat{l}}, \hat{\phi}_{\hat{l}}, \hat{\theta}_{\hat{l}})\odot \big(\bD_{\R,1} \big(\bee_{\R}(\hat{\phi}_{\hat{l}})\odot   {\ba}_{\R,k}(\hat{\phi}_{\hat{l}}) \big) \bX_{\R,2,\hat{l}}^{(u)}[k] \big)} \Big) \notag\\
& \odot {\bf G}_k(\hat{\tau}_{\hat{l}}, \hat{\phi}_{\hat{l}}, \hat{\theta}_{\hat{l}}) \bigg) (\bX_{\R,2,\hat{l}}^{(u)}[k])^T \bigg] \odot \big(\bee_{\R}(\hat{\phi}_{\hat{l}}) \odot {\ba}_{\R,k}(\hat{\phi}_{\hat{l}})\big) \Bigg\}.
\end{align}
\endgroup

\bibliographystyle{IEEEtran}
\bibliography{DLBSrefs2} 

\end{document}